\documentclass[preprint]{aastex63}
\usepackage{mathtools}
\usepackage{color}
\usepackage{multirow}
\usepackage{comment}

\turnoffeditone

\newcommand{\Tabref}[1]{Table~\ref{#1}}

\newcommand{\Equrefs}[1]{Equations~(\ref{#1})}
\newcommand{\Figref}[1]{Figure~\ref{#1}}
\newcommand{\Figrefs}[1]{Figures~\ref{#1}}
\newcommand{\Secref}[1]{Section~\ref{#1}}

\newcommand{\mr}{\mathrm}

\received{June 1, 2021}
\revised{September 27, 2021}
\accepted{September 30, 2021}
\submitjournal{Astronomical Journal}

\shorttitle{Size distribution of small main-belt asteroids}
\shortauthors{Maeda et al.}

\begin{document}

\title{Size distributions of bluish and reddish small main-belt asteroids obtained by Subaru/Hyper Suprime-Cam
\footnote{This research is based on data collected at Subaru Telescope, which is operated by the National Astronomical Observatory of Japan. We are honored and grateful for the opportunity of observing the Universe from Maunakea, which has the cultural, historical and natural significance in Hawaii.}}

\correspondingauthor{Natsuho Maeda}
\email{nmaeda@stu.kobe-u.ac.jp}

\author{Natsuho Maeda}
\affiliation{Department of Planetology, Kobe University, 1-1 Rokkodai-cho, Nada-ku, Kobe 657-8501, Japan}

\author{Tsuyoshi Terai}
\affiliation{Subaru Telescope, National Astronomical Observatory of Japan, National Institutes of Natural Sciences (NINS), 650 North A`ohoku Place, Hilo, HI 96720, USA}

\author{Keiji Ohtsuki}
\affiliation{Department of Planetology, Kobe University, 1-1 Rokkodai-cho, Nada-ku, Kobe 657-8501, Japan}

\author{Fumi Yoshida}
\affiliation{School of Medicine, University of Occupational and Environmental Health, 1-1 Iseigaoka, Yahata, Kitakyusyu 807-8555, Japan}
\affiliation{Planetary Exploration Research Center, Chiba Institute of Technology, 2-17-1 Tsudanuma, Narashino, Chiba 275-0016, Japan}

\author{Kosuke Ishihara} 
\affiliation{Department of Planetology, Kobe University, 1-1 Rokkodai-cho, Nada-ku, Kobe 657-8501, Japan}

\affiliation{Department of Astronomical Science, School of Physical Sciences, Graduate University for Advanced Studies (SOKENDAI), 2-21-1 Osawa, Mitaka, Tokyo 181-8588, Japan}
\affiliation{National Astronomical Observatory of Japan, 2-21-1 Osawa, Mitaka, Tokyo 181-8588, Japan} 

\author{Takuto Deyama}
\affiliation{Department of Planetology, Kobe University, 1-1 Rokkodai-cho, Nada-ku, Kobe 657-8501, Japan}

\affiliation{NTT Data MSE Co., 1-2-33, Miyahara, Yodogawa-ku, Osaka-shi, Osaka 532-0003, Japan}








\begin{abstract}
We performed a wide-field survey observation of small asteroids using the Hyper Suprime-Cam installed on the 8.2 m Subaru Telescope. We detected more than 3,000 main-belt asteroids with a detection limit of 24.2~mag in the $r$-band, which were classified into two groups (bluish C-like and reddish S-like) by the $g-r$ color of each asteroid and obtained size distributions of each group. 
We found that the shapes of size distributions of asteroids with the C-like and S-like colors agree with each other in the size range of $0.4-5$~km in diameter. Assuming the asteroid population in this size range is under collision equilibrium, our results indicate that compositional difference hardly affects the size dependence of impact strength, at least for the size range between several hundred meters and several kilometers. This size range corresponds to the size range of ``spin-barrier'', an upper limit observed in the rotation rate distribution. Our results are consistent with the view that most asteroids in this size range have a rubble-pile structure.
\end{abstract}



\section{INTRODUCTION} \label{sec:intro}

The main asteroid belt, located between the orbits of Mars and Jupiter, consists of asteroids with various spectral types. S-type and C-type are two of the most major types in the main-belt \citep{dc14}.
S-type asteroids are thought to have a rocky composition, and to be the parent bodies of ordinary chondrite meteorites because the abundance and spectral features are consistent with each other \citep[e.g.,][]{b01,b02}. However, the shapes of their spectra do not exactly match and this problem was previously recognized as a missing link between the asteroids and the meteorites. Understanding of space weathering was advanced by laboratory experiments \citep[e.g.,][]{s01} as well as spacecraft studies by NEAR Shoemaker \citep{pr02} and Hayabusa. The Hayabusa spacecraft brought a sample of an S-type asteroid (25143) Itokawa back to the Earth, and analysis of the sample revealed that S-type asteroids are parent bodies of ordinary chondrites \citep{n11}.
On the other hand, C-type asteroids are thought to have a carbon-rich composition with hydrous minerals \citep{u19,po20} and are likely to be parent bodies of carbonaceous chondrites. A sample from a C-type asteroid (162173) Ryugu was returned by the Hayabusa2 spacecraft and is now under analysis. Another sample-return mission by OSIRIS-REx for a C-group (a broader definition of C-type) asteroid (101955) Bennu is also underway. The exact composition of C-type asteroids will soon be revealed.

S-type and C-type asteroids likely have different compositions, and their formation regions should also be different. In the current main-belt, S-type and C-type asteroids dominate in the inner-belt and the middle-/outer-belts, respectively, and their heliocentric distributions partly overlap \citep{dc14, yn07}.
This indicates that asteroids were mixed radially by some process (e.g., planetary migration) in the early Solar System \citep[e.g.,][]{w11,n18,y19,y20}.
Therefore, understanding the evolutions of asteroids with different spectral types can provide a clue to reveal the Solar System's history. In the present work, we focus on the size distribution of asteroids with different spectral types.
Since the current size distributions of asteroids are produced as a result of continuous collisional evolution, comparing the size distributions of asteroids with different spectral types is expected to provide important insights into their collisional evolution.

Many observational studies have been carried out about the size distribution of main-belt asteroids \citep[MBAs; e.g.,][]{jm98,i01,y03,yn04,yn07,w07,p08,g09,t13,aw13,r15,p20}. Some of them studied the size distributions of MBAs with different spectral types. \citet{i01} analyzed the Sloan Digital Sky Survey (SDSS) data and detected about 13,000 asteroids.
They classified their asteroid sample into bluish and reddish groups, based on the principal component analysis using the $g-r$ and $r-i$ colors. They obtained the size distribution for each group in the range of absolute magnitude $10< H < 19$ ($1\lesssim D \lesssim 40$~km, $D$: diameter).
They approximated their cumulative size distribution with a power-law function
\begin{equation}
N(>D) \propto D^{-b},
\end{equation}
where $N(>D)$ is the number of asteroids with diameters larger than $D$.
They found that the power-law indices of the size distributions of bluish/reddish groups agree with each other in the size range of $5<D< 10$ km ($b=4.00\pm 0.05$) while they are slightly different in the size range of $3<D<5$ km ($b=1.40 \pm 0.05$ for the bluish group and $b=1.20\pm 0.05$ for the reddish group).
\citet{yn07} performed a wide-field survey observation with the Suprime-Cam installed on the Subaru Telescope and detected about 1,000 MBAs. They classified their asteroid sample into bluish (C-like) and reddish (S-like) groups based on $B-R$ color, and obtained the size distributions for each group in the range of $14.6 < H < 20.2$. They found that the power-law indices of the bluish and reddish groups agree with each other within errors at $0.3 < D < 1$ km ($b=1.33\pm 0.03$ for bluish asteroids and $b=1.29\pm 0.02$ for reddish asteroids).
These works showed slightly different results, and whether bluish and reddish asteroids have similar size distributions is still inconclusive.

In this study, we performed a wide-field survey observation with the Hyper Suprime-Cam installed on the 8.2 m Subaru Telescope. We detected over 3,000 small MBAs and measured the size distributions of bluish and reddish asteroids using a large homogeneous statistical sample about three times larger than \citet{yn07}, which allowed us to measure more accurate size distributions.
This can provide a clue to better understanding of the collisional evolution of asteroids with different spectral types. 
The structure of this paper is as follows. In \Secref{sec:methods}, we describe our survey observation and data analysis. In \Secref{sec:results}, we show the results of color and size distributions. Comparison with previous works and implications of our results are presented in \Secref{sec:discussion}. We summarize our conclusions in \Secref{sec:conclusions}.

\section{OBSERVATION AND DATA ANALYSIS} \label{sec:methods}
\subsection{Observation}
\begin{figure}[htb]
	\begin{center}
		\includegraphics*[bb=-54 0 666 756,scale=0.4]{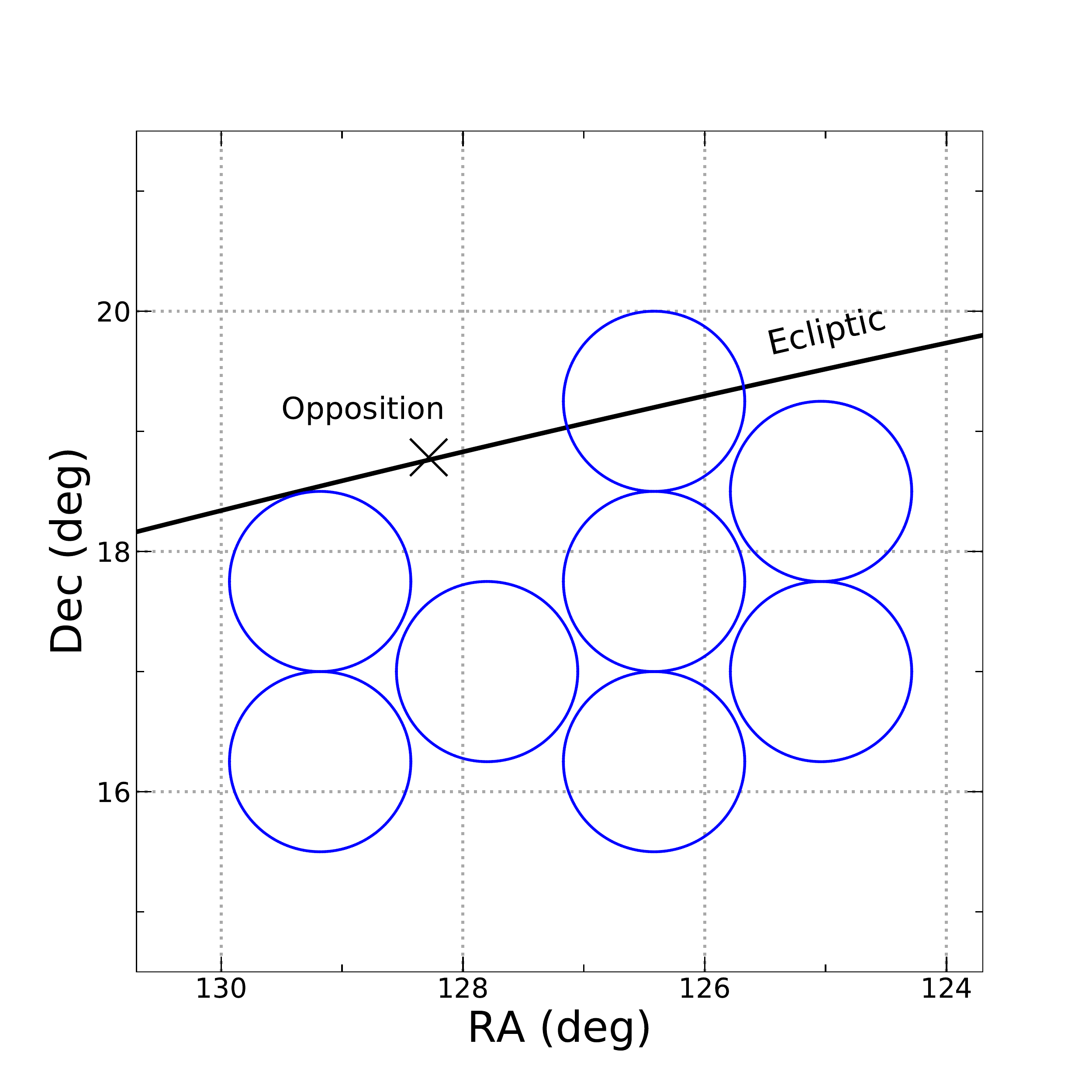}
	\end{center}
	\caption{Locations of the surveyed fields of this work. Each circle represents the area covered by the field-of-view of HSC. The solid line shows the ecliptic plane. The cross shows the opposition on the observation day.}
	\label{fig:obs_FOV}
\end{figure}

We performed the survey observation on January 26, 2015 (UT) using the Hyper Suprime-Cam (HSC) installed on the 8.2 m Subaru Telescope. HSC is a wide-field prime focus camera with 104 CCDs for science (2048 pix $\times$ 4096 pix). The diameter of a field of view is 1.5 degrees with a pixel scale of 0.17 arcseconds \citep{m18}. We used the $g$-band (wavelength $0.40-0.55\;\mr{\mu m}$) and the $r$-band (wavelength $0.55-0.70\;\mr{\mu m}$) filters \citep{k18}.
We obtained the imaging data of eight fields near the opposition and the ecliptic plane, covering the sky area of about $14 \;\mr{deg^2}$. The survey area is shown in \Figref{fig:obs_FOV}.

The observation procedure is as follows:
First, we visited each field five times with the $r$-band at intervals of about 40 minutes. Next, we visited each field in the same way with the $g$-band. We can reduce the effect of brightness variation due to asteroids' rotation by averaging five values of an asteroid's flux obtained with each filter. The time of single exposure is 200 seconds. We summarize our observation in \Tabref{tab:survey_data}.

\begin{table}
\caption{Observation log.}
\begin{center}
\fontsize{9pt}{9pt}\selectfont
\begin{tabular}{ccccccc}
\\ \hline \hline
 Field ID & UT & R.A. & Decl. & Filter & Seeing & Number of \\ 
  && (h:m:s) & (d:m:s) & & (arcsec) & Detection \\ \hline
 FIELD01 & 06:51 - 09:25 & 08:20:09 & +18:30:00 & $r$ & 0.59 - 0.99 & 468 \\ 
 & 11:33 - 14:03 & & & $g$ & 1.03 - 1.48 & \\
 FIELD02 & 06:55 - 09:29 & 08:25:41 & +19:15:00 & $r$ & 0.60 - 0.99 & 398 \\
 & 11:38 - 14:07 & & & $g$ & 0.98 - 1.69 & \\
 FIELD03 & 07:00 - 09:33 & 08:25:41 & +17:45:00 & $r$ & 0.54 - 0.90 & 405 \\
 & 11:43 - 14:11 & & & $g$ & 0.85 - 1.57 & \\
 FIELD04 & 07:05 - 09:37 & 08:36:43 & +17:45:00 & $r$ & 0.58 - 0.78 & 427 \\
 & 11:48 - 14:15 & & & $g$ & 1.02 - 1.55 & \\
 FIELD05 & 07:16 - 09:45 & 08:36:43 & +16:15:00 & $r$ & 0.55 - 0.69 & 439 \\
 & 11:58 - 14:23 & & & $g$ & 0.93 - 1.58 & \\
 FIELD06 & 07:21 - 09:49 & 08:31:12 & +17:00:00 & $r$ & 0.54 - 0.76 & 512 \\
 & 12:03 - 14:27 & & & $g$ & 1.11 - 1.78 & \\
 FIELD07 & 07:26 - 09:53 & 08:25:41 & +16:15:00 & $r$ & 0.59 - 0.95 & 363 \\
 & 12:08 - 14:31 & & & $g$ & 1.04 - 2.84 & \\
 FIELD08 & 07:31 - 09:56 & 08:20:09 & +17:00:00 & $r$ & 0.57 - 0.97 & 447 \\
 & 12:13 - 14:35 & & & $g$ & 1.15 - 1.65 & \\ \hline 
\end{tabular}
\label{tab:survey_data}
\end{center}
\end{table}
\subsection{Image Reduction and Detection}\label{sec:id}
We processed the data into calibrated images and created source catalogs using the hscPipe (ver.4.0.5), the HSC data reduction/analysis pipeline software \citep{bo18}. We used the Pan-STARRS 1 (PS1) $3\pi$ catalog \citep{t12,s12,m13} for astrometric and photometric calibrations. \edit1{The astrometric model in hscPipe includes a ninth-order polynomial distortion that is continuous over the full focal plane composed with distinct translation and rotation transforms for every CCD. For each visit, a different polynomial distortion is used for fitting \citep[for details, see][]{bo18}.}

Our developed software extracted candidates of moving objects from the lists of light sources detected by hscPipe based on the following conditions: 
(i) the sky motion is within the range of $\dot{\lambda}>-48$~arcsec~hr$^{-1}$ and $|\dot{\beta}|<-2.5\dot{\lambda}-55$~arcsec~hr$^{-1}$, where $\dot{\lambda}$ and $\dot{\beta}$ are the motions along ecliptic longitude and latitude, respectively (see gray dot-dashed lines in \Figref{fig:lambda-beta}). 
(ii) the objects are detected from four or more out of five visits in the $r$-band data.
\edit1{Then, we searched each object in each $g$-band image based on the position and velocity measured in the $r$-band images assuming a constant motion on the sky.}
Finally, we checked all of the candidates with eyes to remove false detections. \edit1{False detections are mainly blooming caused by electrons leaking from saturated pixels by a very bright object, or cosmic rays.} The resulting data set contains 3,809 MBA candidates.

\subsection{Measurements}\label{sec:measurements} 
We measured the fluxes of the detected MBAs by aperture photometry using the same technique as in \citet{yt17}.
(i) We correct the centroid positions of each object by fitting an object model generated based on its motion and the seeing size of the image. 
(ii) We measure the flux of each object using a moving circular aperture formed based on the sky motion and the seeing size. If another source such as a field star exists in the aperture, we subtract another image taken at the same field from the original image using the Optimal Image Subtraction method \citep{al98,br08} to measure the accurate flux of the target object.
(iii) We convert the flux to magnitude using the photometric zero point measured by hscPipe.
We defined the value of each object's apparent magnitude for each band as the average of all the visits.
\edit1{We rejected objects for which we could not obtain apparent magnitudes in either $r$- or $g$-band, because we could not obtain $g-r$ color for such objects. Such unsuccessful measurements are mainly caused by the presence of bright stars in some of the apertures and the inability to properly subtract stars.}
Finally, we obtained both $r$- and $g$-band apparent magnitudes for 3,472 MBAs.

An apparent magnitude of an asteroid varies with time due to its rotation if it has a non-axisymmetric shape with respect to its spin axis. We estimated how our survey cadence reduced this effect using a Monte Carlo simulation. According to the previous studies for asteroids' rotation period distribution, average values of the peak-to-peak amplitude and rotation period are 0.4~mag and 5.0~hours, respectively, for asteroids with a diameter of several kilometers \citep{w15,p02}. Using these values, we generated 100,000 synthetic objects having a light curve given by a sinusoidal function with a random initial rotation phase, and performed a virtual observation of these asteroids with conditions similar to our observations. We found that the standard deviations of their apparent magnitudes and colors are 0.034~mag and 0.066~mag, respectively, excluding the photometric uncertainties.

\subsection{Orbit Estimation}\label{sec:OrbEle} 
\begin{figure}[htb]
	\begin{center}
		\includegraphics*[bb=0 0 558 500,scale=0.5]{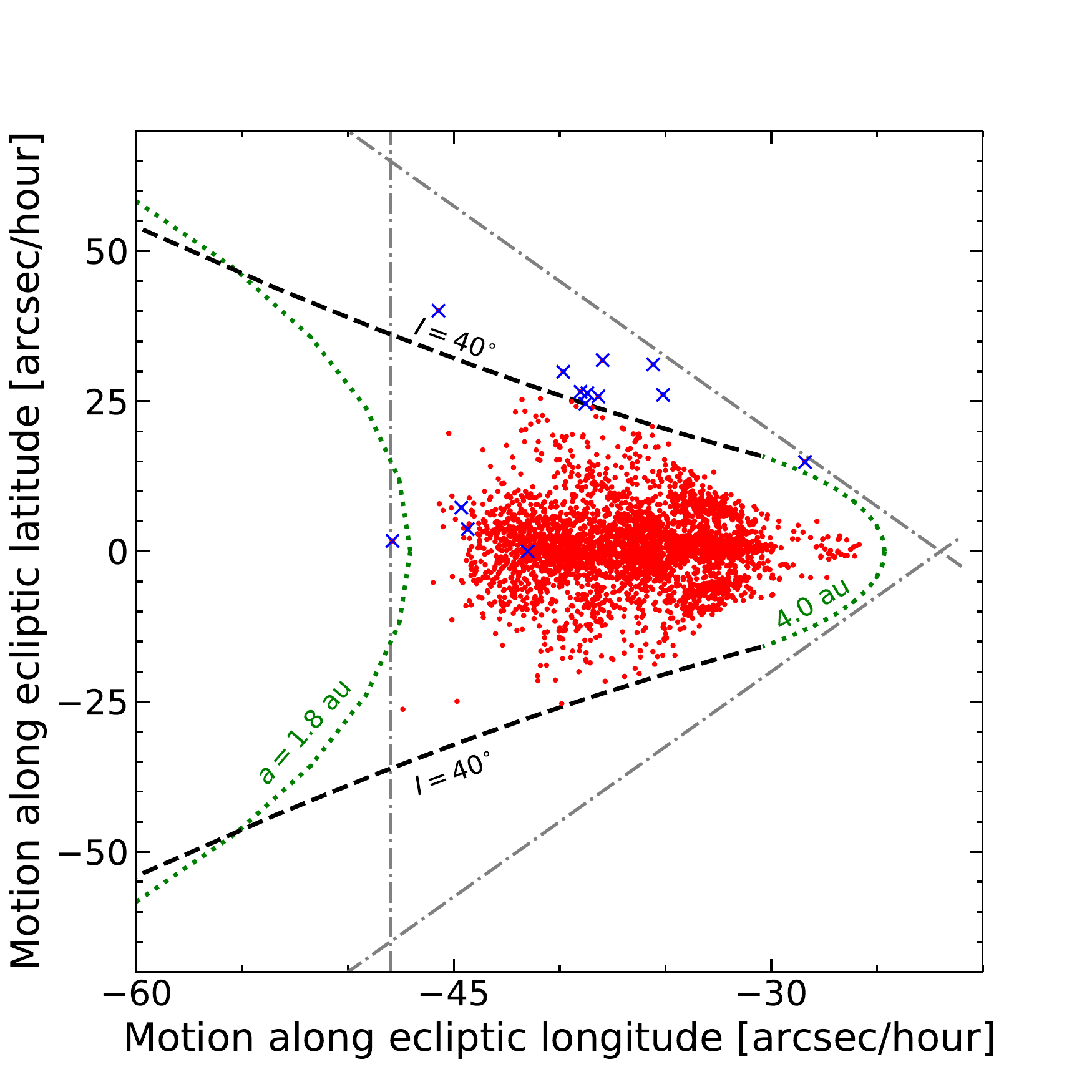}
	\end{center}
	\caption{Distribution of apparent velocities of the detected moving objects. The horizontal axis shows the motion along the ecliptic longitude $\dot{\lambda}$, and the vertical axis shows the motion along the ecliptic latitude $\dot{\beta}$. The objects for which photometric measurements with both $r$- and $g$-bands were successfully performed are plotted. The blue crosses show the objects eliminated from the sample because of large residuals (see in text). The gray dot-dashed lines shows the conditions for MBA detection (\Secref{sec:id}), i.e., $\dot{\lambda}>-48 \; \mr{arcsec\;hr^{-1}}$ and $|\dot{\beta}|<-2.5\dot{\lambda}-55 \; \mr{arcsec\;hr^{-1}}$. The green dotted lines show $a=1.8$~au (left) and $a=4.0$~au (right), respectively, assuming $R=a$ and $e=0$ in \Equrefs{eq:Jedicke1996}. The black dashed lines show $I=40\arcdeg$, $R=a$, and $e=0$ in \Equrefs{eq:Jedicke1996}.}
	\label{fig:lambda-beta}
\end{figure}

\begin{figure}[htb]
\begin{tabular}{cc}

\begin{minipage}{0.5\hsize}
	\begin{center}
		\includegraphics*[bb=0 0 558 500,scale=0.5]{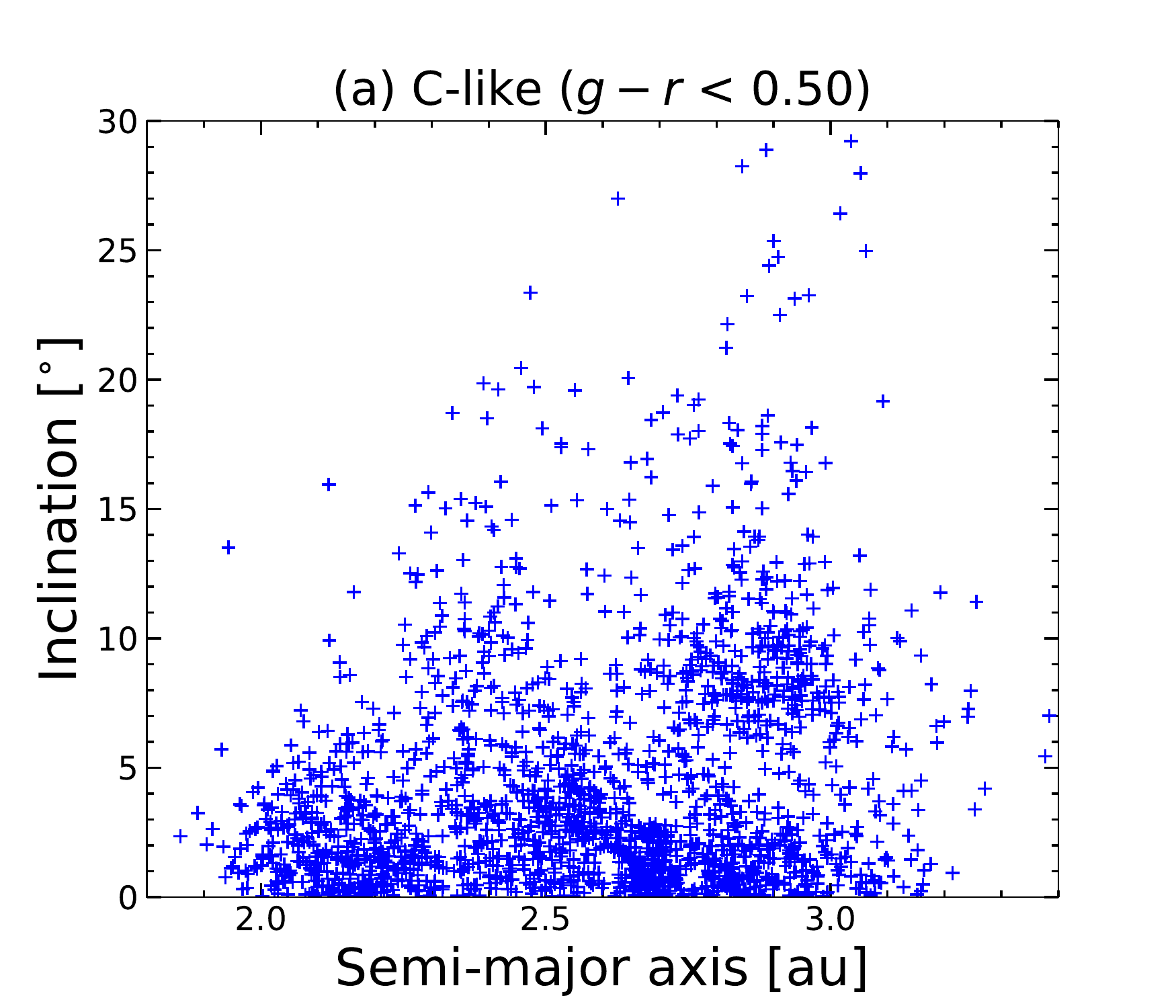}
	\end{center}
\end{minipage}&

\begin{minipage}{0.5\hsize}
	\begin{center}
		\includegraphics*[bb=0 0 558 500,scale=0.5]{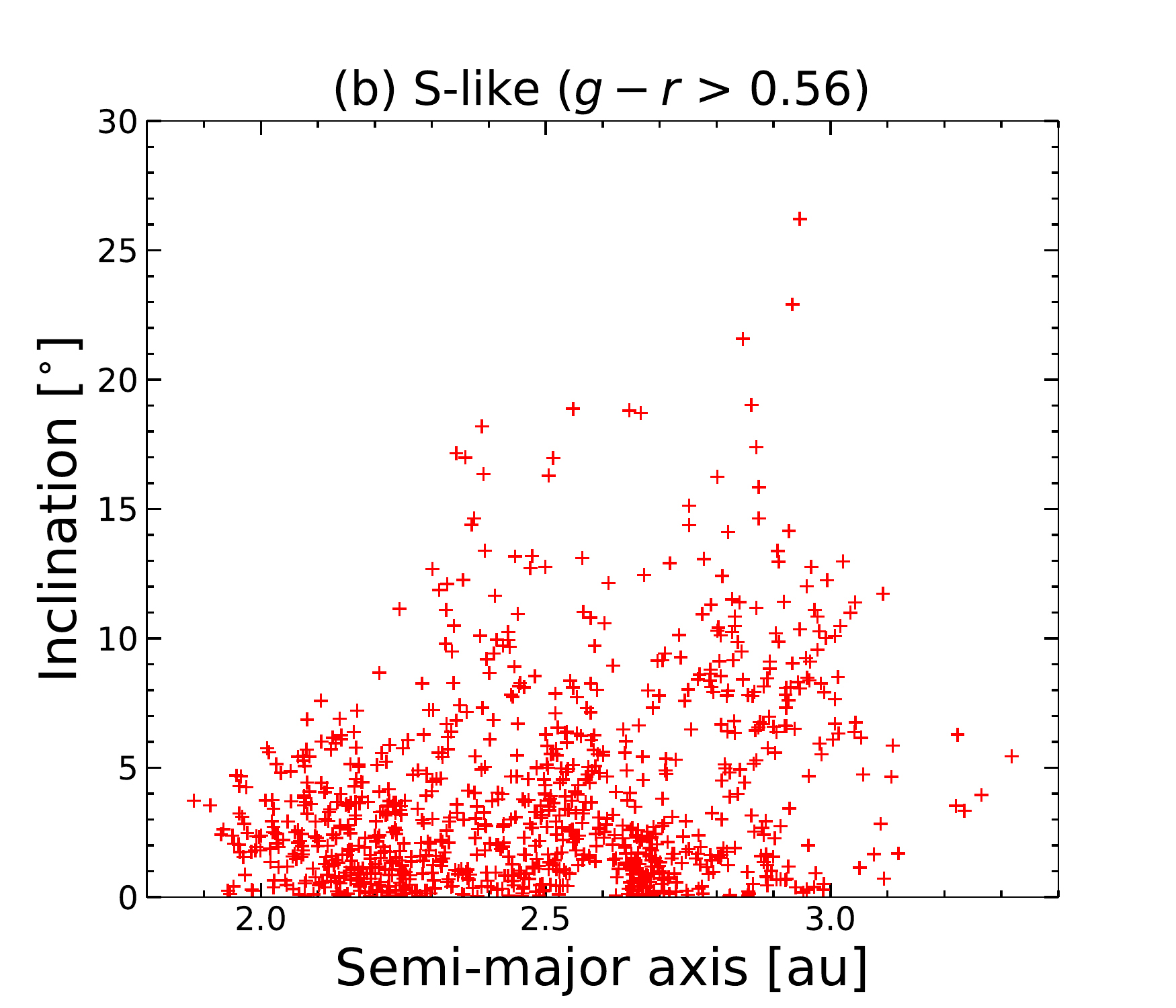}
	\end{center}
\end{minipage}

\end{tabular}
\caption{Semi-major axis vs. inclination of each object in our sample. (a) C-like asteroids ($g-r< 0.50$) and (b) S-like asteroids ($g-r> 0.56$). See \Secref{sec:color-bunrui} for the classification of these two groups.}
	\label{fig:a-i}
\end{figure}

We estimated the orbital elements for each of the detected asteroids from their apparent motion. \citet{j96} derived the relationship between an object's apparent motion velocity in ecliptic coordinates $(\dot{\lambda},\; \dot{\beta})$ and its orbital elements, assuming that they are located near the opposition:
\begin{eqnarray}\label{eq:Jedicke1996}
\begin{dcases}
\dot{\lambda}=\frac{\sqrt{GM_{\odot}}}{R-1}\left( \frac{\sqrt{a(1-e^2)}}{R}\cos I-1 \right),\\
\dot{\beta}=\pm \frac{\sqrt{GM_{\odot}}}{R(R-1)}\sqrt{a(1-e^2)}\sin I,
\end{dcases}
\end{eqnarray}
where $a$, $I$, $e$, and $R$ are object's semi-major axis, inclination, eccentricity, and heliocentric distance, respectively. $G$ and $M_{\odot}$ are the gravitational constant and the mass of the Sun, respectively.
We cannot determine the orbital elements of detected asteroids accurately, because we performed our survey observation only one night. Therefore, we estimated their semi-major axis $a$ and inclination $I$ under the assumptions that they are on circular orbits and were observed at the opposition (i.e., $e=0$, $R=a$, $\lambda^{\prime}=0\degr$, $\beta=0\degr$, where $\lambda^{\prime}$ is ecliptic longitude from the opposition). 
These parameters were determined as the best-fit values for the \Equrefs{eq:Jedicke1996} with the measured sky motion through a least-square method.
If the square root of the sum of squared residuals is larger than 0.01~arcsec~hr$^{-1}$, we remove the object from our sample (we removed 13 asteroids for this reason).
We show the distribution of apparent velocities of 3,472 asteroids in \Figref{fig:lambda-beta}. The removed asteroids are shown with blue crosses.
We obtained a sample of 3,459 MBAs for which $g$- and $r$-band magnitudes as well as orbital elements are successfully obtained. We show the distribution of their orbital elements in \Figref{fig:a-i} (we will explain the classification into C-like/S-like asteroids in \Secref{sec:color-bunrui}).

The statistical errors in the obtained orbital elements were estimated by a Monte Carlo simulation as follows \citep{t13,ny02}:
(i) We generate 100,000 synthetic asteroids randomly distributed in our survey field.
The orbital elements of these asteroids are given with pseudo-random numbers that follow the distribution of orbital elements of known MBAs taken from the ASTORB database \citep{b94}.
(ii) We calculate their apparent motion velocities under the same condition as our survey, and estimate their orbital elements from their apparent motion using \Equrefs{eq:Jedicke1996} under the assumptions of circular orbits and observation near the opposition.
We found that the root mean square errors of estimated $a$, $I$, $R$, and $\Delta$ are $\sigma_a=0.12$~au, $\sigma_I=6.8\arcdeg$, $\sigma_R=0.35$~au, and $\sigma_{\Delta}=0.35$~au, respectively. These values are roughly consistent with the previous works that used the same method \citep{t13,ny02}.

\section{RESULTS} \label{sec:results}
\subsection{Sample Selection}\label{sec:sample_sele} 
We derived the absolute magnitude of each asteroid from its apparent magnitude $m$, heliocentric distance $R$, and geocentric distance $\Delta$ as \citep{b89}
\begin{equation}\label{eq:H}
H = m - 5 \log(R\Delta) - P(\theta),
\end{equation}
where $P(\theta)$ is the phase function at a solar phase angle $\theta$ described as

\begin{equation}
P(\theta)=-2.5\log [(1-G)\Phi_1 + G\Phi_2].
\end{equation}
In the above, $G$ is the slope parameter. We use a typical value of MBAs, $G=0.15$, because there is no way to obtain magnitudes at various solar phase angles in this survey. $\Phi_1$ and $\Phi_2$ are functions of $\theta$ given as
\begin{eqnarray}
\Phi_i=\exp (-A_i [\tan (\theta/2)]^{B_i}) \quad (i=1,2), \nonumber \\
A_1 = 3.33,\; A_2 = 1.87,\; B_1 = 0.63,\; B_2 = 1.22.
\end{eqnarray}

We can describe an error of absolute magnitude $\sigma_H$ considering error propagation as follows:
\begin{equation}
\sigma_{H} = \sqrt{\sigma_{m}^2 + \left( -\frac{5}{\ln 10}\frac{\sigma_R}{R} \right)^2 + \left( -\frac{5}{\ln 10}\frac{\sigma_{\Delta}}{\Delta} \right)^2 },
\end{equation}
where $\sigma_m$ is the photometric error of apparent magnitude. As we described in \Secref{sec:OrbEle}, the errors of heliocentric and geocentric distances are $\sigma_R=\sigma_{\Delta}=0.35$~au. Therefore, we can estimate the error of absolute magnitude to be $\sigma_H\simeq 0.4-0.8$~mag.

\begin{figure}[htb]
\begin{tabular}{cc}

\begin{minipage}{0.5\hsize}
	\begin{center}
		\includegraphics*[bb=0 0 558 500,scale=0.5]{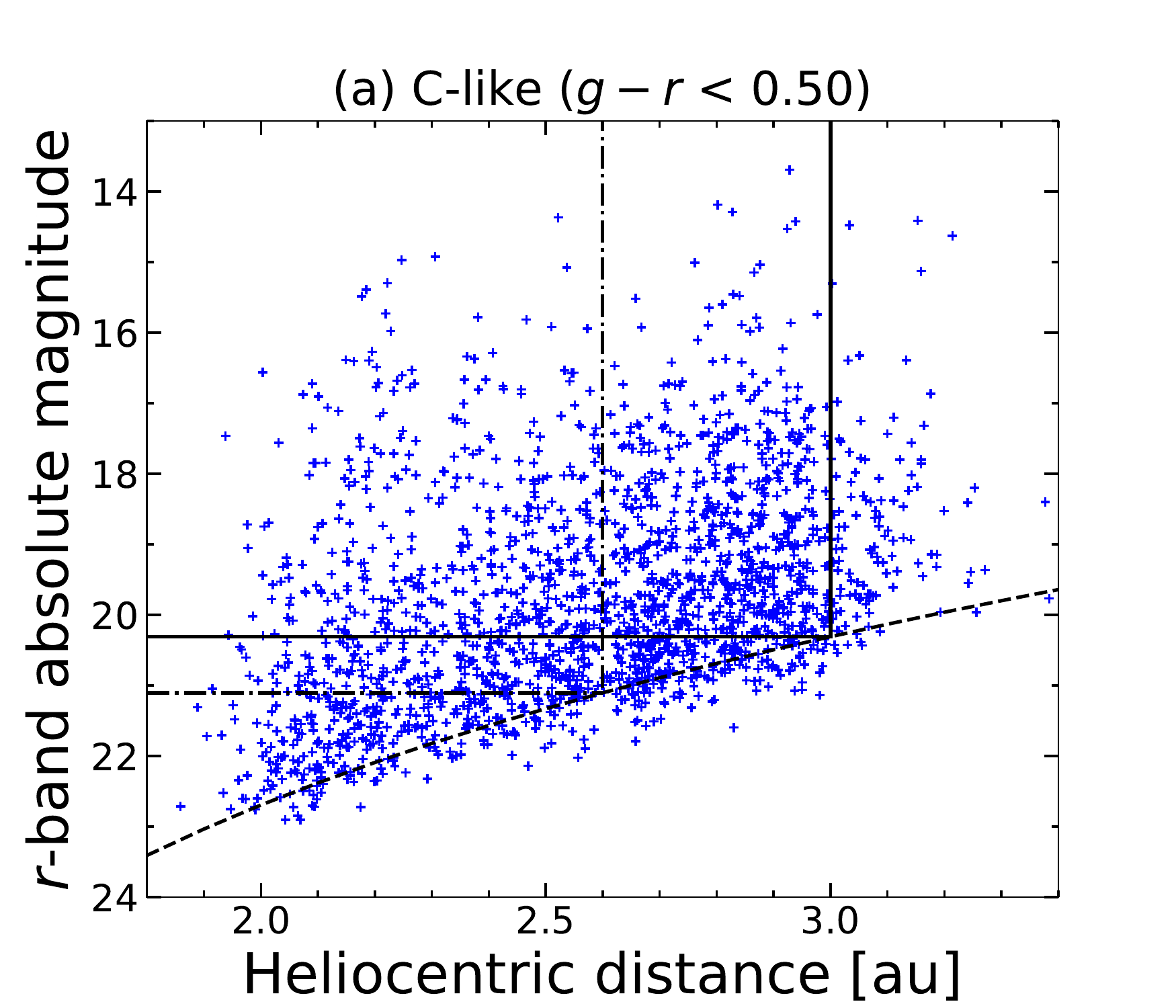}
	\end{center}
\end{minipage}&

\begin{minipage}{0.5\hsize}
	\begin{center}
		\includegraphics*[bb=0 0 558 500,scale=0.5]{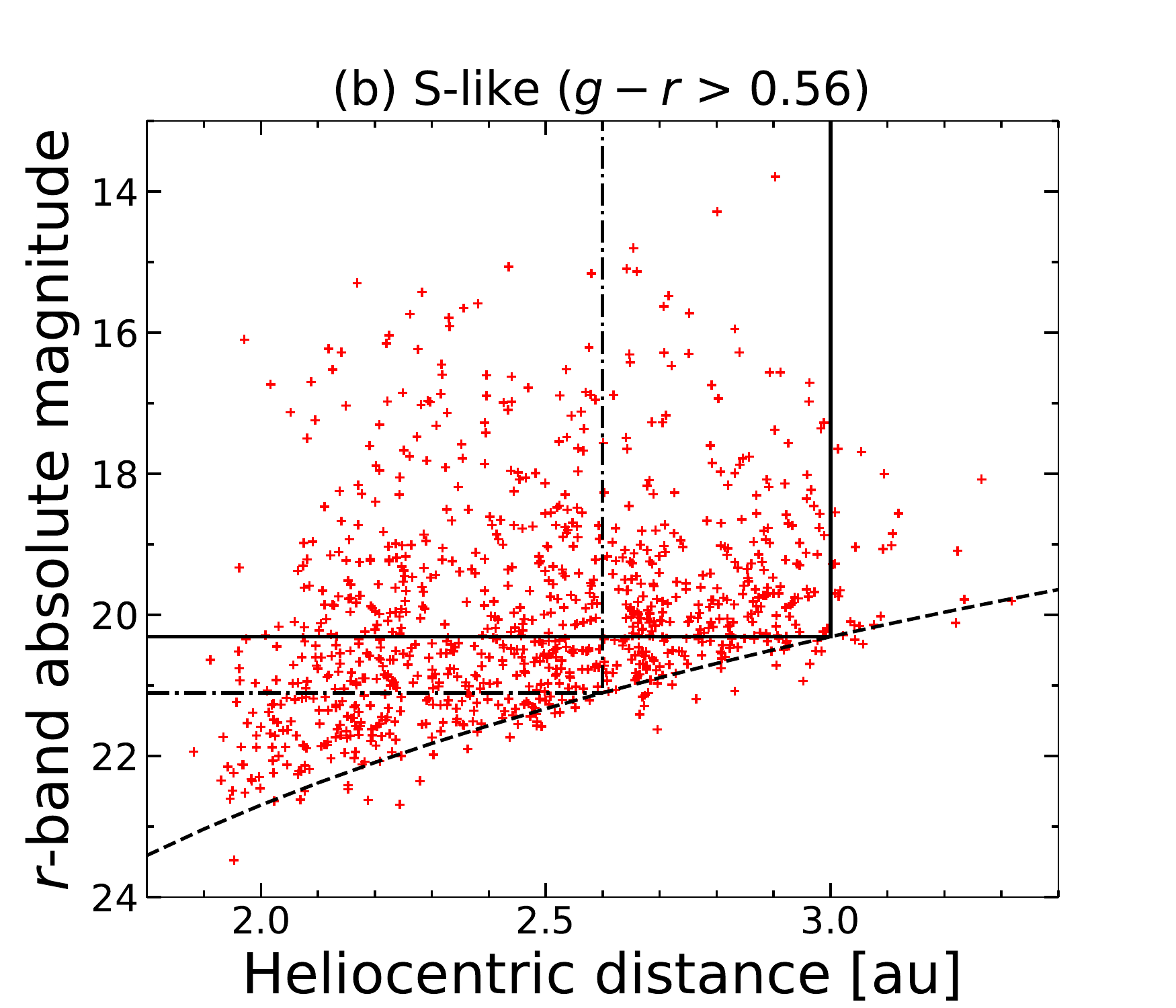}
	\end{center}
\end{minipage}

\end{tabular}
\caption{Heliocentric distance vs. $r$-band absolute magnitude of each object in our sample. (a) C-like asteroids ($g-r< 0.50$) and (b) S-like asteroids ($g-r > 0.56$). See \Secref{sec:color-bunrui} for the classification of these two groups. The dashed line shows apparent magnitude of $m_r=24.2$~mag. The solid lines show the detection limit of the whole region, i.e., heliocentric distance of $R=3.0$~au and absolute magnitude of $H_r=20.3$ mag. The dot-dashed lines show the detection limit for the inner region, i.e., heliocentric distance of $R=2.6$~au and absolute magnitude of $H_r=21.1$~mag.}
	\label{fig:sample_sele}
\end{figure}

To eliminate biases induced by detection incompleteness, we set the detection limit magnitude of our survey based on detection efficiencies.
We examined the detection efficiency through the data reduction procedure with implanted synthetic moving objects \citep[see][for details]{yt17}. Then, the detection efficiency for each CCD image obtained as a function of the apparent magnitude $m$ of implanted objects is fit by the following function
\begin{equation}
\eta(m) = \eta_0 \sum_{k=1,2}\frac{\epsilon_k}{2}\left[ 1-\tanh \left( \frac{m-m_{50}}{w_k} \right) \right]
\end{equation}
with
\begin{equation}
\epsilon_1(m) = 1-\epsilon_2(m) = \frac{1}{2} \left[ 1-\tanh \left( \frac{m-m_{50}}{0.2} \right) \right],
\end{equation}
where $\eta_0$ is the maximum value of the detection efficiency, $m_{50}$ is the apparent magnitude where detection efficiency is $\eta_0/2$, and $w_k\;(k=1,2)$ is the transition width. We calculate the best-fit parameter values of detection efficiencies for all the exposures for each CCD and set 24.2~mag as the detection limit of apparent magnitude where the detection efficiency is 50\% or more for most ($\sim$90\%) of the exposure/CCD images. \Figref{fig:sample_sele} shows the plots of $R$ and $H_r$ for each of the 3,459 objects in our sample (we will explain the classification into C-like/S-like asteroids in \Secref{sec:color-bunrui}). We defined the outer edge of our MBA sample as $R=3.0$~au, where $m_r=24.2$~mag corresponds to $H_r=20.3$~mag. Thus, we select objects with $R \leq 3.0$~au and $H_r \leq 20.3$~mag as an unbiased sample. 
We also investigated absolute magnitude distribution for each region divided by a heliocentric distance 2.6~au, i.e., inner region ($R\leq 2.6$~au) and outer region ($2.6\leq R \leq 3.0$~au), respectively. The detection limit for the inner region corresponds to $H_r=21.1$~mag. Note that the above choice of the boundaries of the inner and outer regions does not greatly affect our main results (\Secref{sec:hd}).

\subsection{Color Distributions}\label{sec:color} 
\begin{figure}[htb]
	\begin{center}
		\includegraphics*[bb=0 0 600 500,scale=0.5]{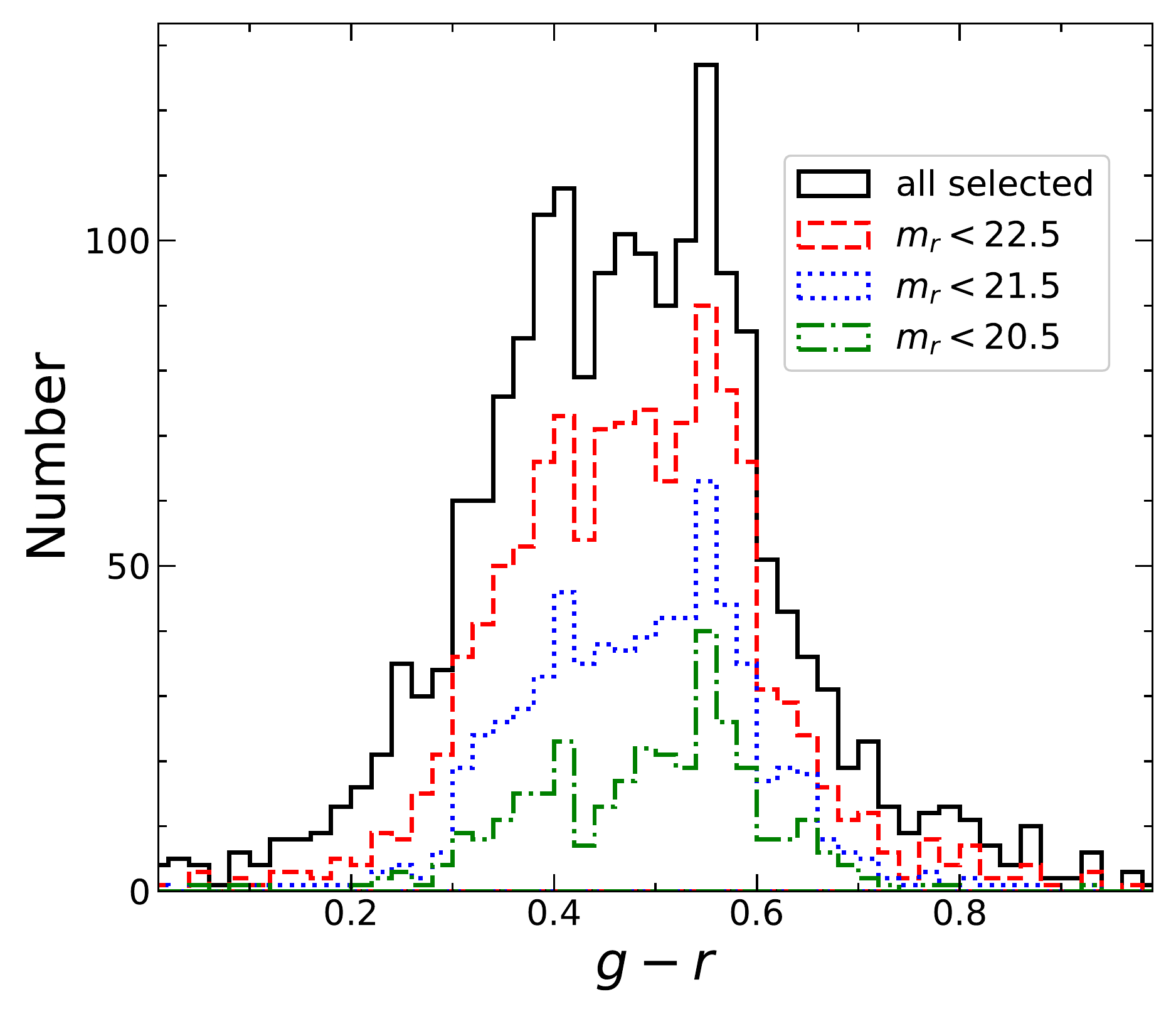}
	\end{center}
	\caption{Histogram of the $g-r$ colors of our unbiased sample. Different lines represent the cases limited by different apparent magnitudes: black for all the selected sample, red for $m_r<22.5$~mag, blue for $m_r<21.5$~mag, and green for $m_r<20.5$~mag.}
	\label{fig:colorhist_limit}
\end{figure}

\Figref{fig:colorhist_limit} shows the histogram of the $g-r$ color of our unbiased sample (\Secref{sec:sample_sele}). In addition to the case of the whole sample, we also plotted the cases limited by various apparent magnitudes $m_r$. When the sample is limited to those with $m_r<21.5$~mag (blue dotted line) or $m_r<20.5$~mag (green dot-dashed line), \edit1{a structure that may be regarded as double peaks is seen in the color distribution.} The peaks are located at $g-r\sim 0.4$ and 0.6, respectively, which roughly correspond to those of the C-type and S-type asteroids, respectively, seen in the 4th release of SDSS Moving Object Catalog \citep[SDSS MOC4,][see \Secref{sec:color-bunrui}, \Figref{fig:bunrui}]{i10,h11}.
\edit1{Although the double peaks are expected to be hidden considering the color uncertainties induced by photometry and asteroids' rotation, two populations can be recognized when the photometric error is low enough ($m_r\lesssim21.5$~mag).\footnote{We tested the visibility of the bimodality of color distribution and found that the double peak cannot be seen when a color uncertainty is assumed to be 0.066~mag as large as the rotational effect of our sample.}
Also, other effects such as space weathering are likely to have relatively minor contributions to the shape of the color distribution\footnote{Little effect of space weathering is also understood by comparison of time-scales between space weathering \citep[less than 1~Myr;][]{v09} and collisional life time \citep[$\sim 10-50$~Myr for MBAs with $0.1-1$~km in diameter;][]{db07}. }. }

Some previous works reported a bimodal color distribution of MBAs, while others found no such dichotomy.
\citet{i01} found bimodal distributions in the color-color diagrams plotted for objects whose photometric errors are less than 0.05~mag, while they showed that the color dichotomy becomes unclear for fainter objects as we found in \Figref{fig:colorhist_limit}. \citet{yn07} reported bimodality of $B-R$ color distribution of sub-km to km-sized MBAs, although it was unclear for the sample for the middle and outer main-belt.
\citet{g09} found no color dichotomy even though their sample has a small error of absolute magnitudes (the error values are 0.11~au and 0.36~mag in heliocentric distance and absolute magnitude, respectively) due to accurate orbits obtained by multiple-night observations. 
\citet{p20} measured $g'-r'$ colors of about 1,000 MBAs with orbits determined and found no significant color dichotomy, which might be caused by uncertainty in absolute magnitudes due to asteroid rotation.

\subsection{Color Classification}\label{sec:color-bunrui} 
\begin{figure}[htb]
	\begin{center}
		\includegraphics*[bb=0 0 576 500,scale=0.5]{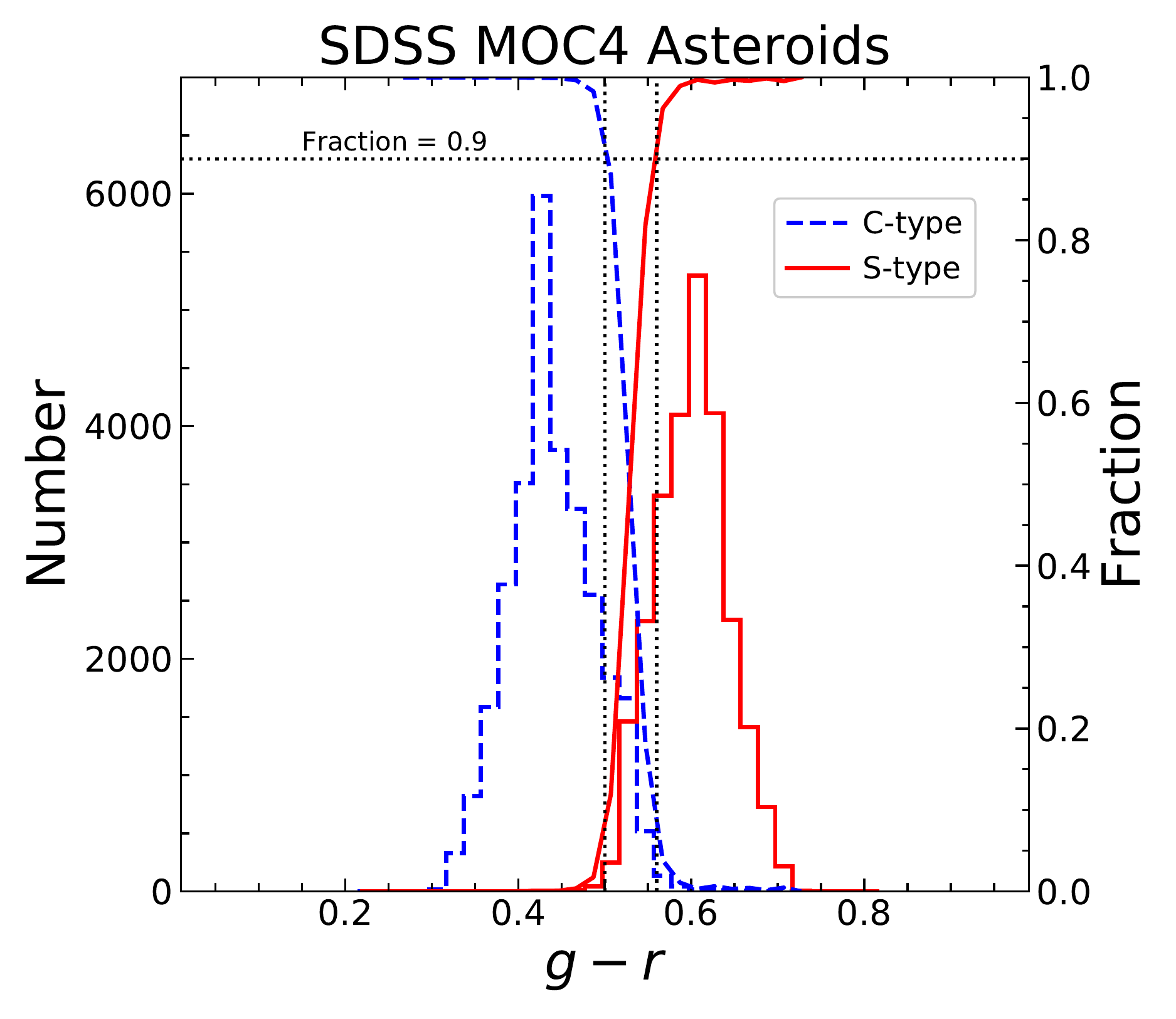}
	\end{center}
	\caption{Histogram of the $g-r$ colors for the SDSS MOC4 \citep{i10,h11} $\mr{C_{p}}$-type (blue dashed-lines) and $\mr{S_{p}}$-type (red solid-lines) asteroids corrected for the difference of the filter systems between HSC and SDSS. The classification for each spectral type is based on the \citet{c10} taxonomy. The curves show the fraction of asteroids belonging to each type in each color bin. The horizontal dotted line shows the fraction of 0.9, and the vertical dotted lines show the color boundaries used for our classification into C-/S-like asteroids, corresponding to the $g-r$ colors at the fraction of 0.9 for each type.}
	\label{fig:bunrui}
\end{figure}

Using the $g-r$ color of each asteroid, we classified our sample into two groups, i.e., bluish C-like asteroids and reddish S-like asteroids.
In order to define the classification criterion, we use the SDSS MOC4. We corrected the color of each asteroid in the SDSS MOC4 to that in HSC's filter system, and classified these objects into spectral types according to the taxonomy given by \citet{c10} and \citet{h11} ($\mr{C_p,\; S_p}$, etc.).
\Figref{fig:bunrui} shows the $g-r$ color histogram of $\mr{C_p}$-/$\mr{S_p}$-type asteroids in the SDSS MOC4. The curves show the fraction of asteroids belonging to each of the two types for each color bin; for example, the blue curve shows (the number of $\mr{C_p}$-type asteroids)/(the total number of $\mr{C_p}$-type and $\mr{S_p}$-type asteroids) for each bin. We defined the classification criteria by the $g-r$ color values where the fraction is 0.9; those with $g-r<0.50$ are defined as C-like asteroids and those with $g-r>0.56$ as S-like asteroids.
In this way, we can get less-contaminated samples than the previous works, \citet{i01} and \citet{yn07}, which used a single color boundary between bluish and reddish groups.

\edit1{We evaluated the accuracy of our classification using the SDSS MOC4 asteroids. We classified $\mr{C_{p}}$-type and $\mr{S_{p}}$-type of the SDSS MOC4 asteroids into C-like and S-like by our classification criteria, respectively. \Tabref{tab:cm} is the confusion matrix of the color classification using the SDSS MOC4 asteroids. We found that the recall of $\mr{C_p}$-type (the fraction of asteroids classified as C-like among all actual $\mr{C_{p}}$-type asteroids) is 99.1\%, and the recall of $\mr{S_{p}}$-type (the fraction of asteroids classified as S-like among all actual $\mr{S_{p}}$-type asteroids) is 99.3\%. We also found that the precision of C-like (the fraction of actual $\mr{C_{p}}$-type asteroids among all asteroids classified as C-like) is 99.4\%, and the precision of S-like (the fraction of actual $\mr{S_{p}}$-type asteroids among all asteroids classified as S-like) is 98.9\%. The $\mr{F_1}$-scores of C-like and S-like are 99.3\% and 99.1\%, respectively.}

\begin{table}[h]
	\begin{center}
	\caption{Confusion matrix of the color classification using the SDSS MOC4 asteroids.}
	\label{tab:cm}
	\begin{tabular}{cc|cc|c}
	& \multicolumn{4}{c}{Classified} \\
\multirow{4}{*}{Actual} & & C-like & S-like & Recall \\ \cline{2-5}
& $\mr{C_{p}}$-type & 25472 & 236 & 99.1\% \\
& $\mr{S_{p}}$-type & 148 & 21602 & 99.3\% \\ \cline{2-5}
& Precision & 99.4\% & 98.9\% &
	\end{tabular}
	\end{center}
	\tablecomments{The rows show the actual types ($\mr{C_{p}}$-type or $\mr{S_{p}}$-type) of classified asteroids, and the columns show their classified types (C-like or S-like).}
	
\end{table}

It should be noted that each of the above two groups likely contains asteroids with other spectral types because we use only the $g-r$ color. C-like asteroids likely contain D and X-types as well as their complexes. Also, S-like asteroids likely contain L, Q, V, D, X, A-types and their complexes.

\subsection{Comparison of Size Distributions of C-like and S-like Asteroids}\label{sec:csd_C-S}
\begin{figure}[htb]
\begin{tabular}{cc}

\begin{minipage}{0.5\hsize}
	\begin{center}
		\includegraphics*[bb=0 0 522 500,scale=0.5]{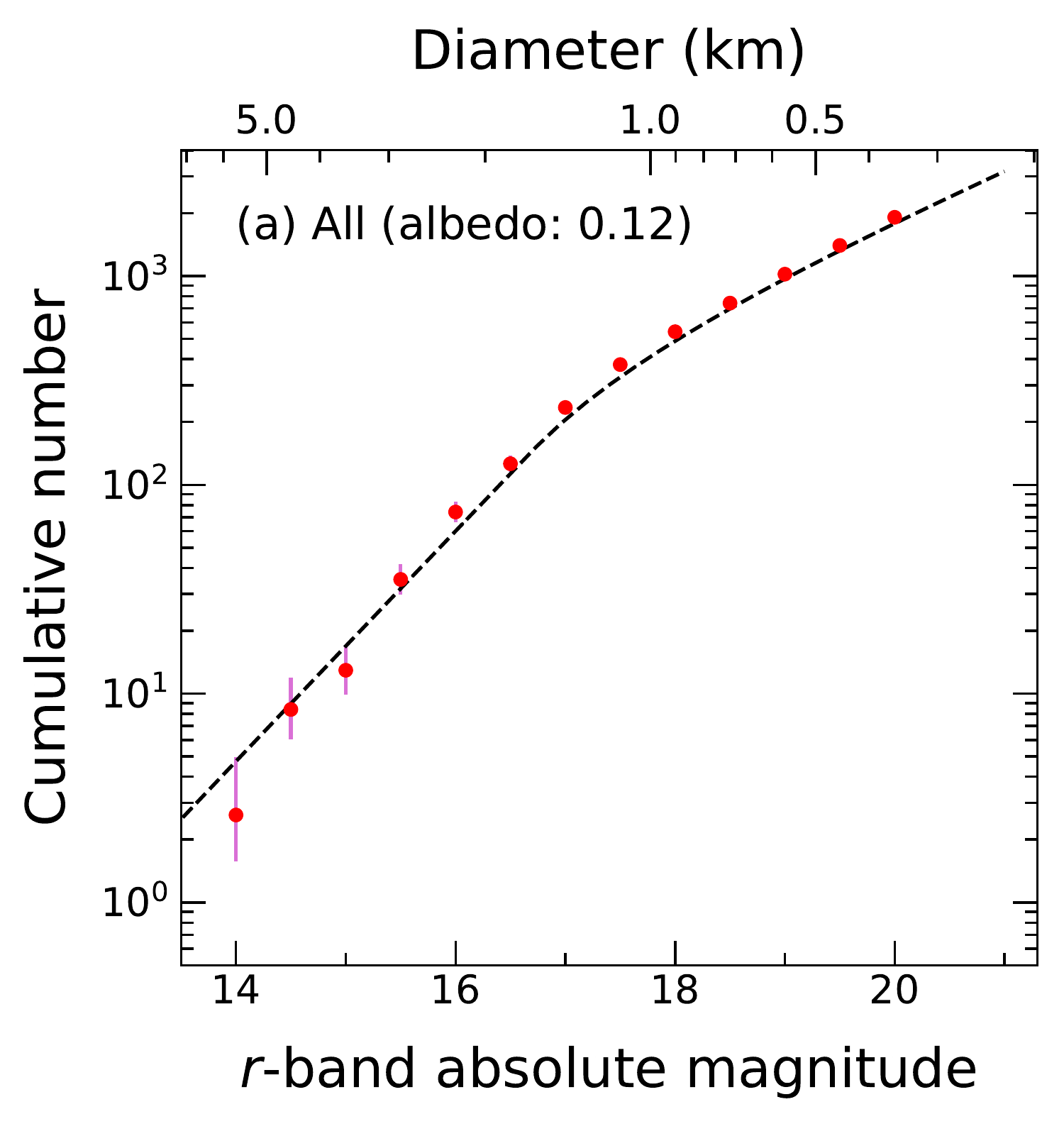}
	\end{center}
	\end{minipage}&
	\\

\begin{minipage}{0.5\hsize}
	\begin{center}
		\includegraphics*[bb=0 0 522 500,scale=0.5]{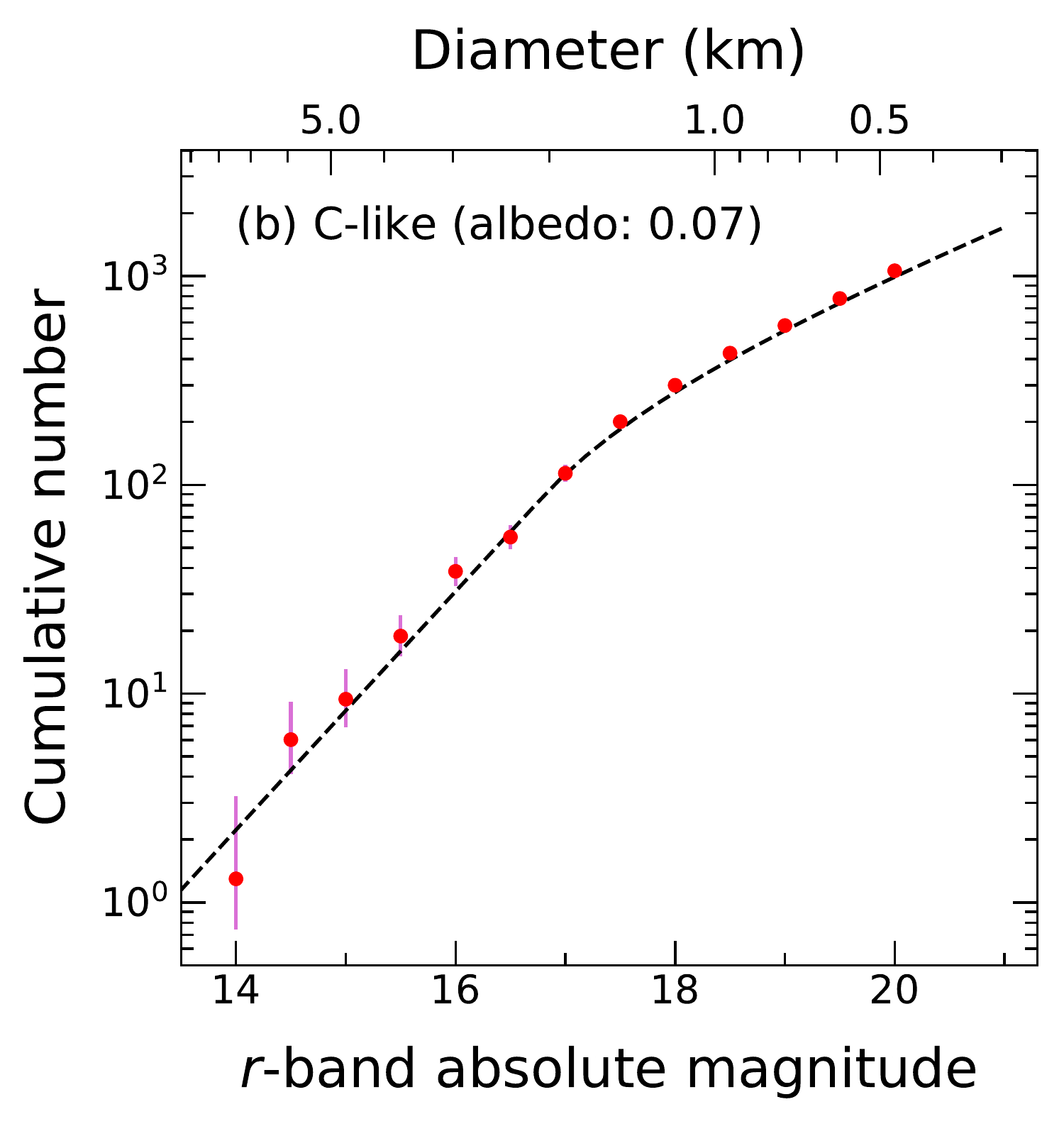}
	\end{center}
\end{minipage}&

\begin{minipage}{0.5\hsize}
	\begin{center}
		\includegraphics*[bb=0 0 522 500,scale=0.5]{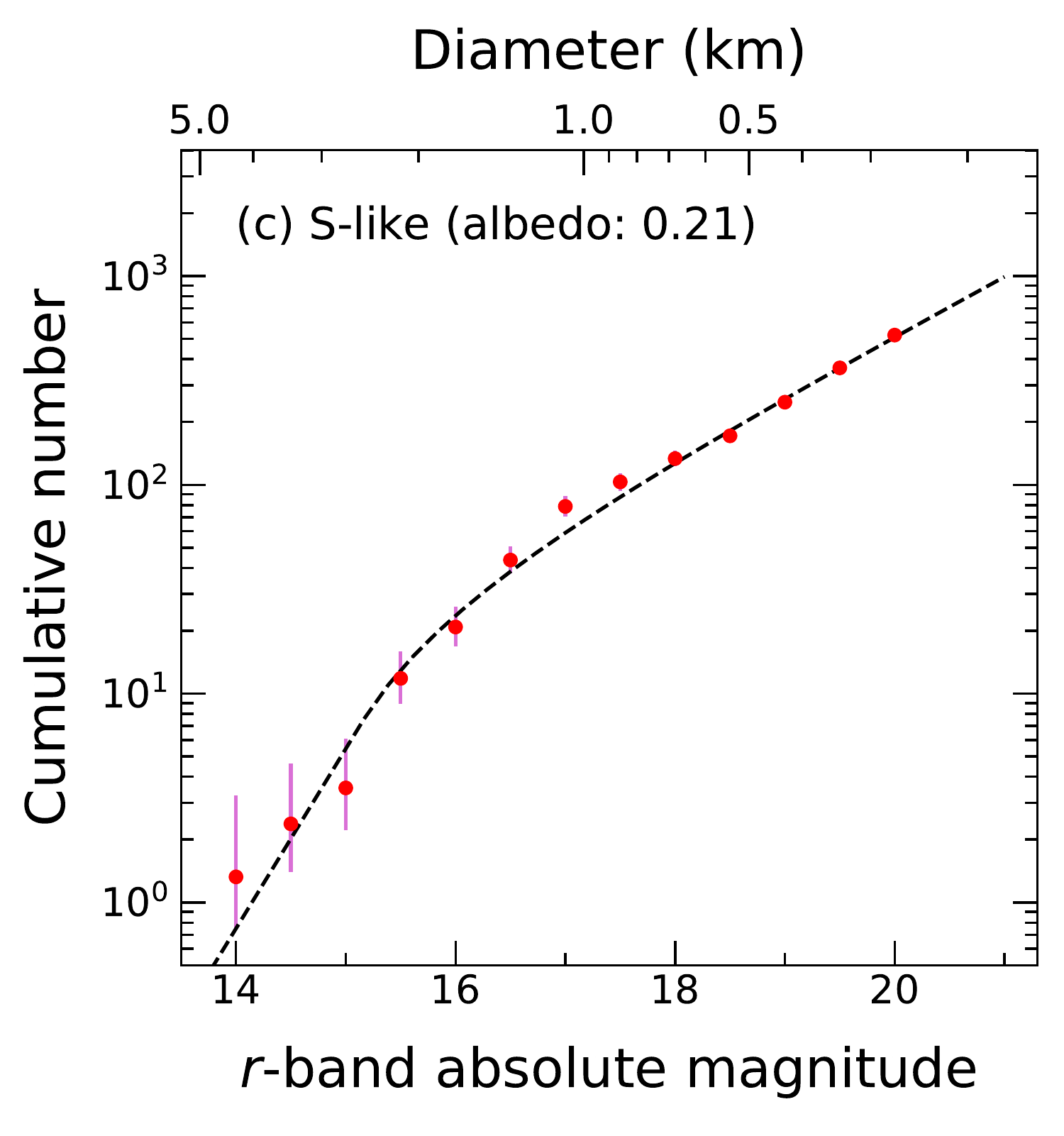}
	\end{center}
\end{minipage}

\end{tabular}
	\caption{Distributions of the $r$-band absolute magnitudes of the objects in our unbiased sample for the whole region. (a) all asteroids, (b) C-like asteroids ($g-r< 0.50$), and (c) S-like asteroids ($g-r> 0.56$), respectively. The red dots show cumulative number of the asteroids for each bin. The bin width is 0.5 mag. The error bars represent Poisson's statistical error. The dashed lines show the best-fit models (see \Tabref{tab:csd}). Diameters shown on the upper horizontal axis were obtained by converting from absolute magnitudes under the assumption of a constant albedo of (a) $p=0.12$, (b) $p=0.07$, and (c) $p=0.21$, respectively (see \Secref{sec:csd_C-S}).}
	\label{fig:csd}
\end{figure}

Based on the photometric measurements and color classification described above, we obtained absolute magnitude distributions for our MBA sample. \Figrefs{fig:csd}(a), (b), and (c) show the absolute magnitude distributions for all, C-like, and S-like asteroids, respectively. The cumulative numbers shown on the vertical axis are corrected by the detection efficiency for each asteroid \citep{yt17}:
\begin{equation}
N(<H)=\sum_{j:H_j<H}\frac{1}{\eta(m_j)},
\end{equation}
where $\eta(m_j)$ is the detection efficiency of the $j$-th object with apparent magnitude $m_j$ in the $r$-band. As described in \Secref{sec:id}, the sample objects were detected from four or five $r$-band images. In this case, the detection efficiency of an asteroid in all the four images taken for a certain field is given by \citep{yt17}:
\begin{equation}
\eta(m) = \prod_{i=1}^4 \eta_i(m).
\end{equation}
where $\eta_i(m)$ is the detection efficiency of this object in the $i$-th image.
We estimated $\eta(m)$ as a product of the largest four values of $\eta_i$ for each object.

In order to compare the size distributions of C-like and S-like asteroids, we convert the absolute magnitude of an asteroid to its diameter using the following equation:
\begin{equation}\label{eq:convert_D}
\log D = 0.2r_{\odot}-\log \frac{\sqrt{p}}{2R_{\oplus}} -0.2H_r,
\end{equation}
where $r_{\odot}=-26.91$ is the $r$-band magnitude of the Sun \citep{f11}, $p$ is a geometric albedo, and $R_{\oplus}$ is the heliocentric distance of the Earth (i.e., 1~au). We assumed the constant albedo of $p=0.07$ and $p=0.21$ for C- and S-like asteroids, respectively, based on the mean values of the C- and S-type MBAs measured by the infrared astronomical satellite AKARI \citep{u13}.

\begin{figure}[htb]
\begin{tabular}{cc}
	
	\begin{minipage}{0.5\hsize}
	\begin{center}
		\includegraphics*[bb=0 0 400 400,scale=0.6]{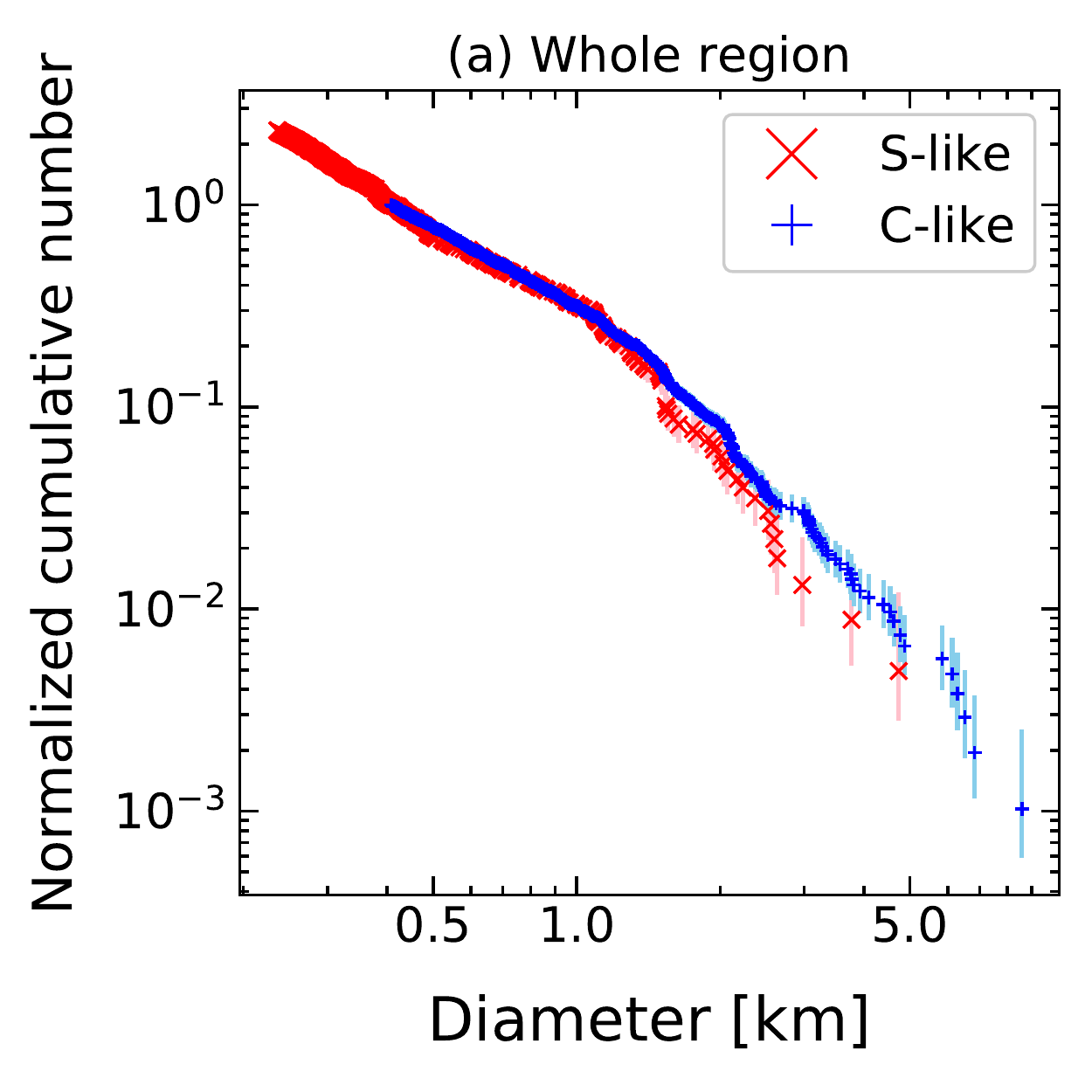}
	\end{center}
	\end{minipage}\\
	
	\begin{minipage}{0.5\hsize}
	\begin{center}
		\includegraphics*[bb=0 0 400 400,scale=0.6]{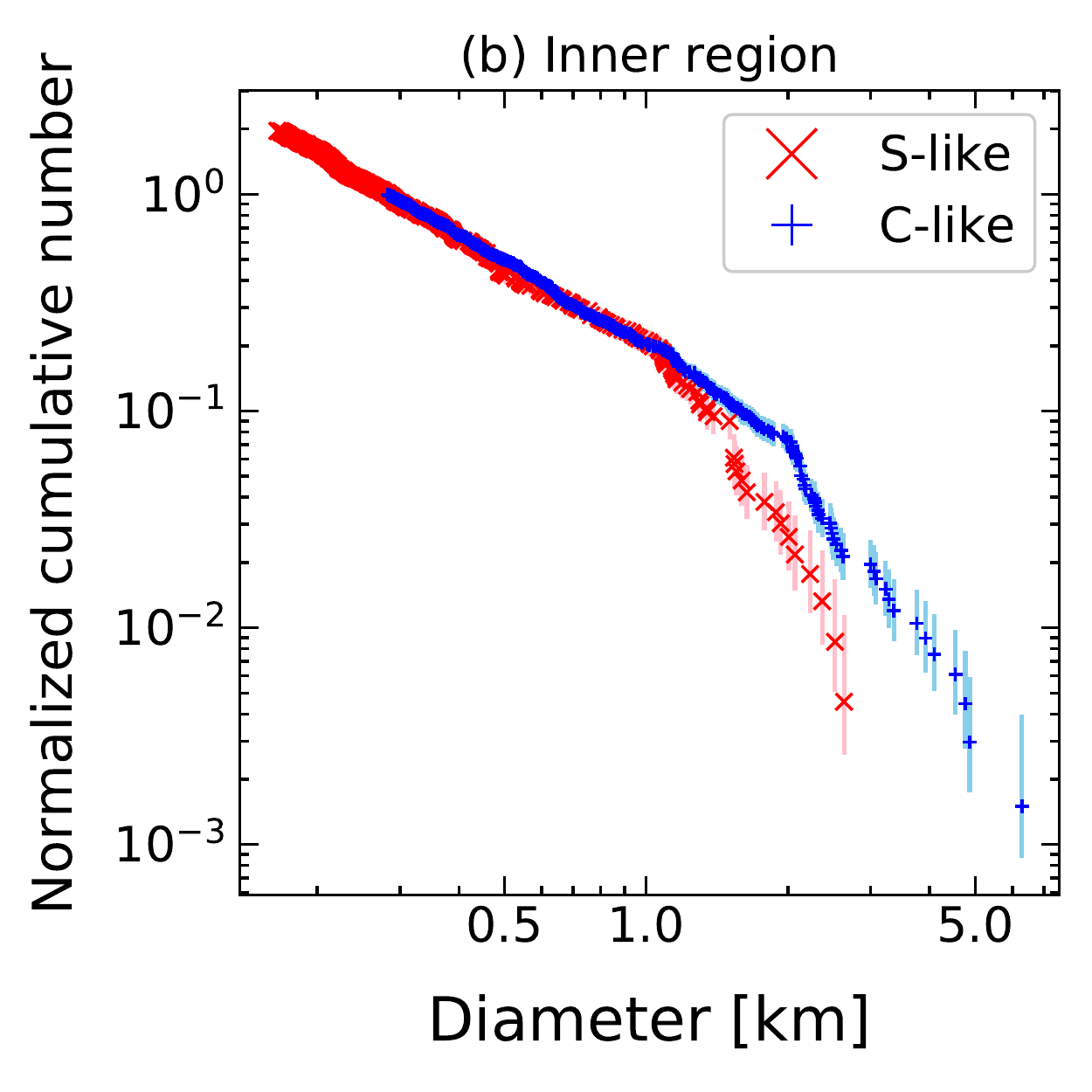}
	\end{center}
	\end{minipage}&
	
	\begin{minipage}{0.5\hsize}
	\begin{center}
		\includegraphics*[bb=0 0 400 400,scale=0.6]{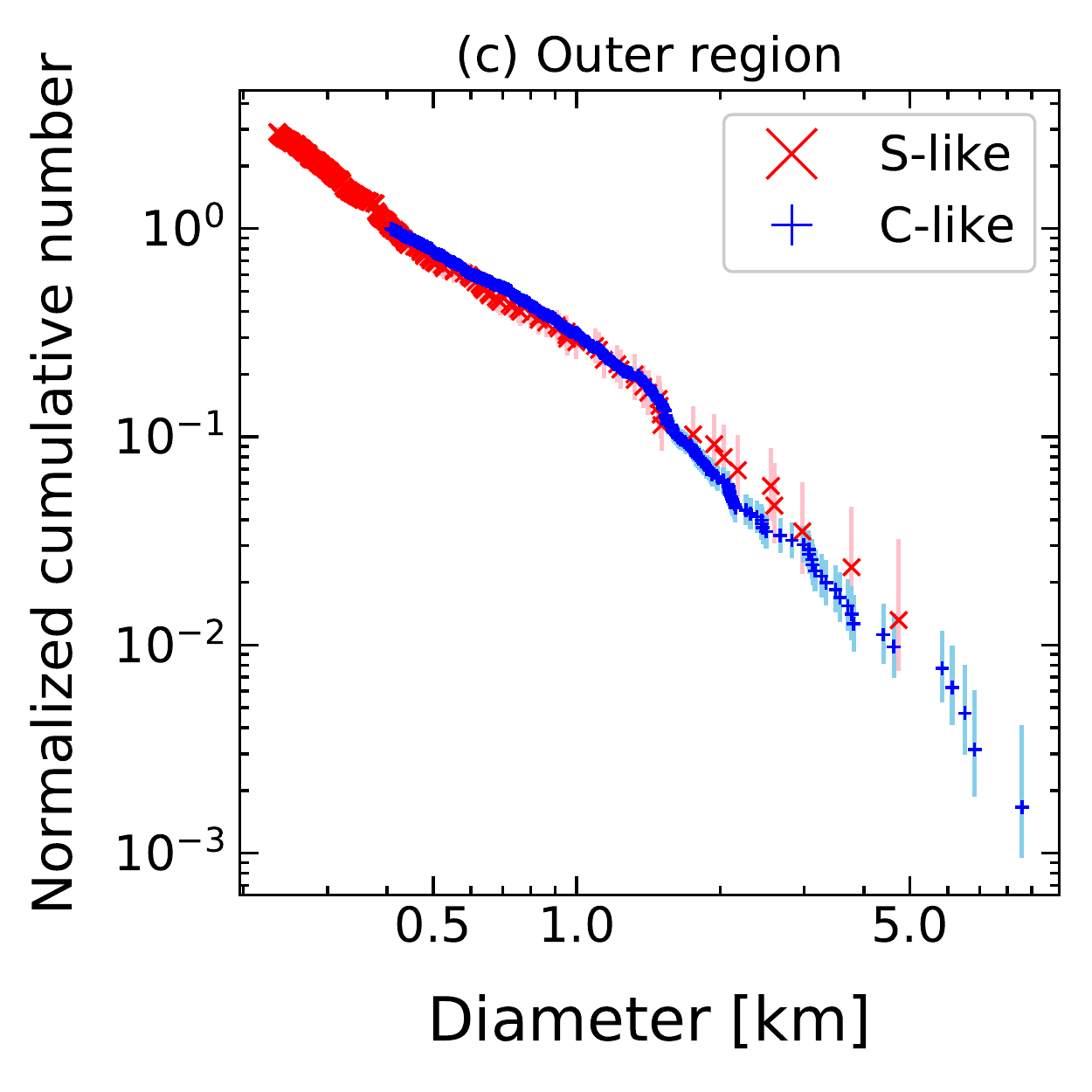}
	\end{center}
	\end{minipage}
	
\end{tabular}
	\caption{Cumulative size distributions of bluish C-like asteroids ($g-r<0.50$; shown by blue pluses) and reddish S-like asteroids ($g-r>0.56$; shown by red crosses). (a), (b), and (c) are for the whole region, the inner region ($1.8<a<2.6$~au), and the outer region ($2.6<a<3.0$~au), respectively. The values of the vertical axis are normalized by the values at $D=0.4$~km. Diameters are obtained by converting from absolute magnitudes under the assumption of a constant albedo of $p=0.07$ for C-like asteroids, and $p=0.21$ for S-like asteroids \citep{u13}, respectively.}
	\label{fig:csd_C-S}
\end{figure}

\Figref{fig:csd_C-S} shows a direct comparison of the cumulative size distributions between the C- and S-like asteroids. The cumulative number in the vertical axis is normalized by the value at $D=0.4$~km, which corresponds to the detection limit of the C-like asteroids. The error bars show Poisson's statistical error. We found that the shapes of the cumulative size distributions of C- and S-like samples agree with each other within statistical errors for $0.4< D < 1.5$~km.

We evaluated the goodness of the agreement by the Kolmogorov-Smirnov test \citep[hereafter KS test,][]{p92}. The statistic $D_{\mr{KS}}$ of the KS test is defined by the maximum deviation between two cumulative distribution functions $S_{1}(x)$ and $S_{2}(x)$ as
\begin{equation}
D_{\mr{KS}}=\mr{max}|S_{1}(x)-S_{2}(x)|.
\end{equation}
The significance probability is approximately calculated by the following function:
\begin{equation}
P(D_{\mr{KS}}>\lambda)=Q_{\mr{KS}}(\sqrt{n_1n_2/(n_1+n_2)}D_{\mr{KS}})
\end{equation}
with
\begin{equation}
Q_{\mr{KS}}(\lambda) = 2\sum ^{\infty}_{j=1} (-1)^{j-1}e^{-2j^2\lambda ^2},
\end{equation}
where $n_1$ and $n_2$ are numbers of data points of $S_{1}(x)$ and $S_{2}(x)$, respectively.

We performed the KS test under the null hypothesis that the two samples are drawn from the same distribution in the size range of $0.4< D < 4.7$~km. We found that the significance probability is 0.218, i.e., the null hypothesis cannot be rejected at a 20\% significance level.
We confirmed that even when we set the classification criterion of C- and S-like samples more strictly (for example, $g-r< 0.40$ for the C-like asteroids, and $g-r> 0.60$ for the S-like asteroids), the significance probability is no less than 0.2.

\begin{figure}[htb]
\begin{tabular}{cc}

\begin{minipage}{0.5\hsize}
	\begin{center}
		\includegraphics*[bb=0 0 500 500,scale=0.55]{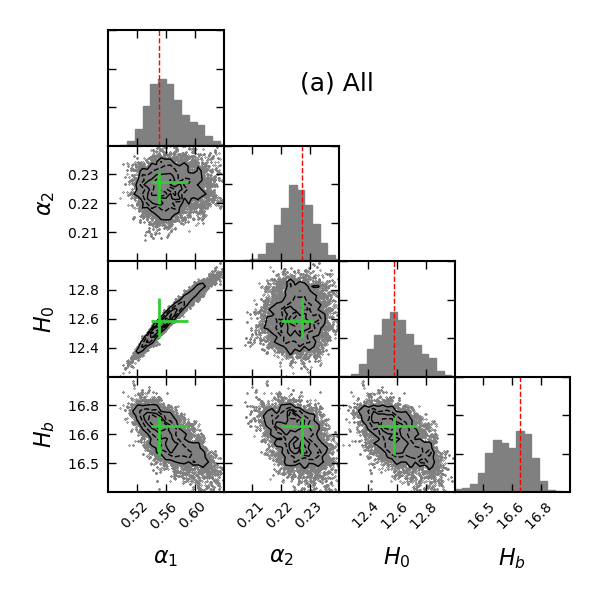}
	\end{center}
	\end{minipage}&
	\\
\begin{minipage}{0.5\hsize}
	\begin{center}
		\includegraphics*[bb=0 0 500 500,scale=0.55]{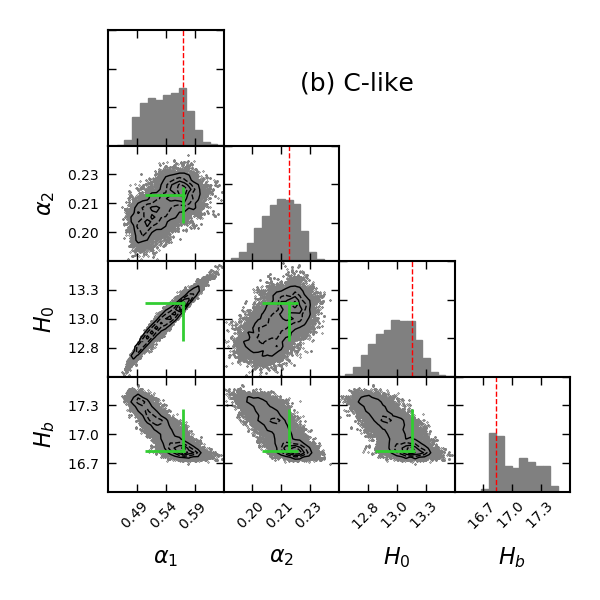}
	\end{center}
\end{minipage}&
\begin{minipage}{0.5\hsize}
	\begin{center}
		\includegraphics*[bb=0 0 500 500,scale=0.55]{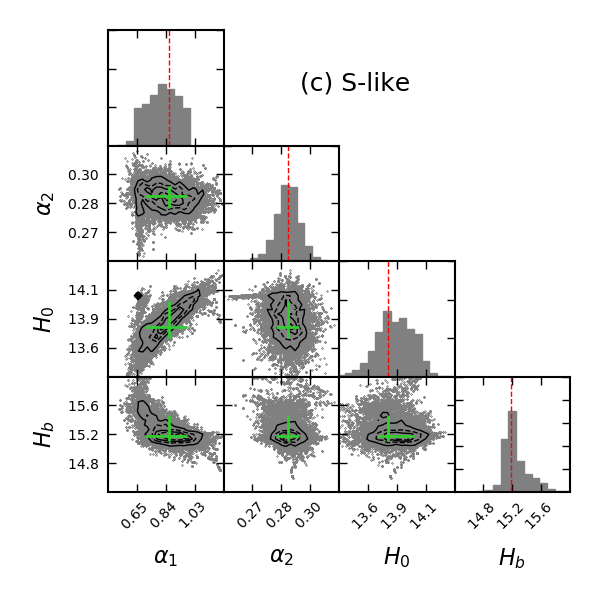}
	\end{center}
\end{minipage}
\end{tabular}
	\caption{Post-event distributions of sampling by MCMC for the whole region. The red dashed lines in the 1D histograms show the best-fit values. Each gray point in the 2D histograms (black contours) shows the result of sampling for each step. The green crosses in the 2D histograms show the error range of the best-fit.}
	\label{fig:corner}
\end{figure}

To measure the slopes of obtained absolute magnitude distributions, we approximated the differential absolute magnitude distributions $\Sigma(H)=dN(H)/dH$ with the following broken power-law:
\begin{equation}\label{eq:bpl}
\Sigma(H)=
	\begin{dcases}
		10^{\alpha_1(H-H_0)} &(H<H_{\mr{break}}), \\
		10^{\alpha_2H +(\alpha_1-\alpha_2)H_{\mr{break}}- \alpha_1H_0} &(H\geq H_{\mr{break}}),
	\end{dcases}
\end{equation}
where $\alpha_1$ and $\alpha_2$ are the power-law indices of the absolute magnitude distributions for the brighter and fainter ends, respectively. $H_{\mr{break}}$ is the absolute magnitude of the breakpoint, and $H_0$ is defined so that $\Sigma(H_0)=1$. We used the maximum-likelihood method \citep{b04,yt17} to obtain the best-fit parameters of \Equrefs{eq:bpl}. For the numerical analysis of the maximum-likelihood estimation, we used the Markov Chain Monte Carlo method (MCMC) with a 
Python package \verb+emcee+\footnote{https://emcee.readthedocs.io/en/stable/} \citep{f13}.
\Figref{fig:corner} shows the post-event distributions of sampling for each parameter by MCMC for the whole region. We adopted \edit1{a maximum a posteriori (MAP)} as the best-fit value for each parameter.

The best-fit parameters are shown in \Tabref{tab:csd}. In all the cases of the all, C-like, and S-like asteroids, the slope $\alpha$ changes at $H \sim 15-17$ mag and the slopes at the fainter end are shallower than those at the brighter end. The slopes in the fainter end are $\alpha_2=0.21^{+0.01}_{-0.01}$ and $\alpha_2=0.28^{+0.01}_{-0.01}$ for C-/S-like asteroids, respectively. On the other hand, the slopes in the brighter end are $\alpha_1=0.57^{+0.01}_{-0.07}$ and $\alpha_1=0.86^{+0.11}_{-0.15}$ for C-/S-like asteroids, respectively. Note that the slopes in the brighter end have large uncertainties due to the small sample number, which was partly caused by CCD saturation for bright objects.

\begin{table}
\caption{Numbers of asteroids in our sample and best-fit parameters for the absolute magnitude distributions.}
\fontsize{11pt}{17pt}\selectfont
\begin{tabular}{cccccc}
\hline \hline
Types & number & $\alpha_1$ & $\alpha_2$ & $H_0$ & $H_{\mr{break}}$ \\ \hline
All & 1814 & $0.55^{+0.04}_{-0.01}$ & $0.23^{+0.01}_{-0.01}$ & $12.58^{+0.17}_{-0.12}$ & $16.69^{+0.04}_{-0.13}$ \\
C-like & 1026 & $0.57^{+0.01}_{-0.07}$ & $0.21^{+0.01}_{-0.01}$ & $13.18^{+0.02}_{-0.33}$ & $16.80^{+0.46}_{-0.01}$ \\
S-like & 479 & $0.86^{+0.11}_{-0.15}$ & $0.28^{+0.01}_{-0.01}$ & $13.80^{+0.20}_{-0.09}$ & $15.14^{+0.29}_{-0.06}$ \\ \hline
\end{tabular}	
	\label{tab:csd}
\end{table}

In \Figref{fig:csd} (c) for S-like asteroids, we notice a significant deviation of the model from the data at $H\sim17$~mag, while they agree well with each other for the cases of all and C-like asteroids. We confirmed that the absolute magnitude distribution for S-like asteroids cannot be approximated well with \Equrefs{eq:bpl} due to the partial excess at $H\sim17$~mag, even after fine-tuning of the parameters. 

In \Figref{fig:csd_C-S} (a), the fainter-end slopes of C-like and S-like asteroids agree with each other, but the best-fit values of $\alpha_2$ are slightly different between them. One reason for this is the discrepancy between the model and the data described above. Another reason is underestimating the number of brighter S-like asteroids mainly because of saturation, which makes the best-fit distribution of S-like asteroids steeper than the actual distribution. 
To accurately evaluate the fainter-end slope, we fit a broken power-law to the S-like asteroid sample again assuming the same values of $\alpha_1$ (0.54) and the diameter at the break corresponding to $H_{\mr{break}}=15.79$ as the best-fit parameters for the C-like asteroid sample.
We found that the best-fit value is $\alpha_2=0.234 \pm 0.002$, which is quite similar to that of the C-like asteroids.

\subsection{Heliocentric Distance Dependence}\label{sec:hd}
We also compared the size distributions for the C- and S-like asteroids in each of the inner and outer regions, separately (\Figrefs{fig:csd_C-S} (b) and (c)).
For the inner region ($1.8< a< 2.6$~au), we found that the shapes of the size distributions of C- and S-like asteroids agree with each other within the statistical errors at $0.3< D < 1.0$~km. The discrepancy in $D>1.0$~km may be caused by undetected bright S-like asteroids due to saturation.
\edit1{The saturation magnitude of our survey is around $m_r = 17.5$~mag, which corresponds to $H_r \simeq 14.5 - 16$~mag ($D=1.7 - 3.4$~km for S-like asteroids) in the inner region. Therefore, underestimation due to saturation is likely to appear for $D\gtrsim 1.7$~km for S-like asteroids ($p=0.21$) in the inner region\footnote{For C-like asteroids ($p \sim 0.07$), the saturation effect is likely to appear for $D\gtrsim 3.4$~km in the inner region.}.}
For the outer region ($2.6<a<3.0$~au), we found that the shapes of the size distributions of C- and S-like asteroids agree with each other within the errors at $0.4< D < 4.5$~km.
We performed the KS test under the null hypothesis that the two samples are drawn from the same distribution in the size ranges of $0.3<D<2.6$~km and $0.4<D<4.7$~km for the inner and outer region, respectively.
We found that the significance probabilities are 0.94 for the inner region and 0.36 for the outer region, respectively, i.e., the null hypothesis cannot be rejected at a 20\% significance level for both regions.
\edit1{For the inner region, we also performed the Anderson-Darling test \citep[][]{ad52}, which is corrected for effects of edges of distributions, and found that the significance probability is 0.107, and the null hypothesis cannot be rejected at a 10\% significance level.}
We also found that their shapes agree with each other within the errors, at least in the size range of $0.4< D < 1.0$~km, even when we change the heliocentric boundary between the two regions by about 0.1~au.
\edit1{It should be noted that the uncertainty of semi-major axis for our sample is $\sigma_a=0.12$~au (\Secref{sec:OrbEle}), thus we have to keep this in mind when discussing the difference between different radial locations.}

\section{DISCUSSION} \label{sec:discussion}
\subsection{Comparison with Previous Works}\label{sec:comp_prevwork}
\citet{yn07} measured size distributions of C-like ($B-R<1.1$) and S-like ($B-R>1.1$) asteroids obtained by a survey observation with the Suprime-Cam on the Subaru Telescope. They found that the slopes for C- and S-like asteroids agree with each other within errors in the fainter end but disagree in the brighter end. The cumulative size distribution of their C-like sample was approximated by a single power-law with the slope of $b=1.33 \pm 0.03$ for $0.6<D<6.5$~km ($\alpha=0.27\pm0.01$ for $14.6<H<20.2$~mag), and that of their S-like sample was approximated by a broken power-law with the slopes of $b_1=2.44 \pm 0.09$ for $1.0<D<2.3$~km ($\alpha_1=0.49\pm0.02$ for $15.4<H<17.0$~mag) and $b_2=1.29 \pm 0.02$ for $0.3<D<1.0$~km ($\alpha_2=0.26$ for $17.4<H<20.2$~mag), respectively, where $b_1$ and $b_2$ are brighter/fainter end slopes of a cumulative size distribution represented by a broken power-law (they have the relationship $b_1=5\alpha_1$ and $b_2=5\alpha_2$, respectively, with the corresponding absolute magnitude distribution slopes $\alpha_1$ and $\alpha_2$).
\citet{i01} measured the size distributions of blue (C-like) and red (S-like) asteroids, which are obtained by SDSS and classified based on the principal component analysis using $g-r$ and $r-i$ colors. They found that the slopes of cumulative size distributions for the brighter ends are $b_1=4.00\pm0.05$ ($\alpha_1=0.80\pm0.01$) for both the blue and the red (the size range was $5\lesssim D \lesssim 20$~km for the blue, and $5\lesssim D \lesssim 40$~km for the red), and that for the fainter ends are $b_2=1.40\pm 0.05$ ($\alpha_2=0.28\pm0.01$) for the blue for $3\lesssim D \lesssim 5$~km, and $b_2=1.20\pm 0.05$ ($\alpha_2=0.24\pm0.01$) for the red for $0.7\lesssim D \lesssim 5$~km \citep[see][for a discussion of the limiting diameters of their size distributions]{y19}.
\citet{p20} measured $g'$- and $r'$-band absolute magnitude distributions for $14<H_{g'}<18$~mag ($1<D<10$~km) with 1,182 MBAs detected from HiTS 2015 data. They did not find any color dependence in the derived absolute magnitude distributions.

These studies showed that both of the C-like and S-like asteroids have similar size distributions with a power-law index of $b\simeq 1.2-1.4$ ($\alpha \simeq 0.24-0.28$) for diameters smaller than several kilometers, but their results have a slight discrepancy.
We confirmed that the size distribution of C- and S-like asteroids have similar shapes at sub-km, and both of them have a break point at $D\sim 2$~km.
From the results of \citet{i01}, \citet{p20}, and the present work, the similarity of the shapes of the size distributions between C-like and S-like asteroids seems to hold up to $\sim$10~km in diameter.

\subsection{Implications}
As we discussed above, the shapes of the size distributions of C- and S-like asteroids are similar in the size range from several hundred meters up to about 10~km.
This seems to indicate a similarity in the size-dependence of the impact strength for the C-/S-like asteroids if asteroid population in this size range is under collision equilibrium.
It has been analytically shown that one of the important factors for determining the shape of a size distribution in collision equilibrium is the size dependence of the impact strength (the specific energy needed to disperse half of the target mass $Q_D^*$) and is insensitive to the size distributions of fragments produced in individual impacts \citep[][]{d69,t96,og03,kt10}.
Therefore, if asteroid population in this size range is under collision equilibrium, the similarity of their size distributions seems to indicate the common size-dependence of $Q_D^*$, which depends on material, internal structure, and impact velocity \citep{ba99,j10}. C-type and S-type asteroids are thought to have different bulk compositions \citep[e.g.,][]{n11,u19,po20,k21}. Considering the above, our results seem to indicate that bulk compositional difference hardly affects the size dependence of $Q_D^*$, at least in size range from several hundred meters to ten kilometers.

On the other hand, the distribution of asteroids rotation periods shows that most asteroids at $0.3\lesssim D \lesssim 10$~km have a rotation period longer than 2.2 hours, the so-called ``spin-barrier'' \citep{p02,c15,c17}. This is the centrifugal disruption limit of asteroids accumulated by only self-gravity. This is often interpreted as indicating that most asteroids in this size range are rubble-piles. The similarity of the size distributions between the C-like and S-like asteroids found in the present work seems to support the view that these asteroids in this size range have rubble-pile structures.

It should be noted that rotational disruption induced by YORP effect is not negligible at $D\lesssim 10$~km \citep[][]{j14}. Also, removal by Yarkovsky effect somewhat affects the size distribution, while collisional disruption primary determine the size distribution of MBAs \citep[][]{og05}. 
Similarity of the shapes of the size distributions of C- and S-like asteroids may indicate that these secondary effect hardly provide difference in size distributions for different spectral types.

\subsection{Rough Estimate of the Total Number of MBAs}
\begin{figure}[htb]

	\begin{center}
		\includegraphics*[bb=0 0 558 500,scale=0.5]{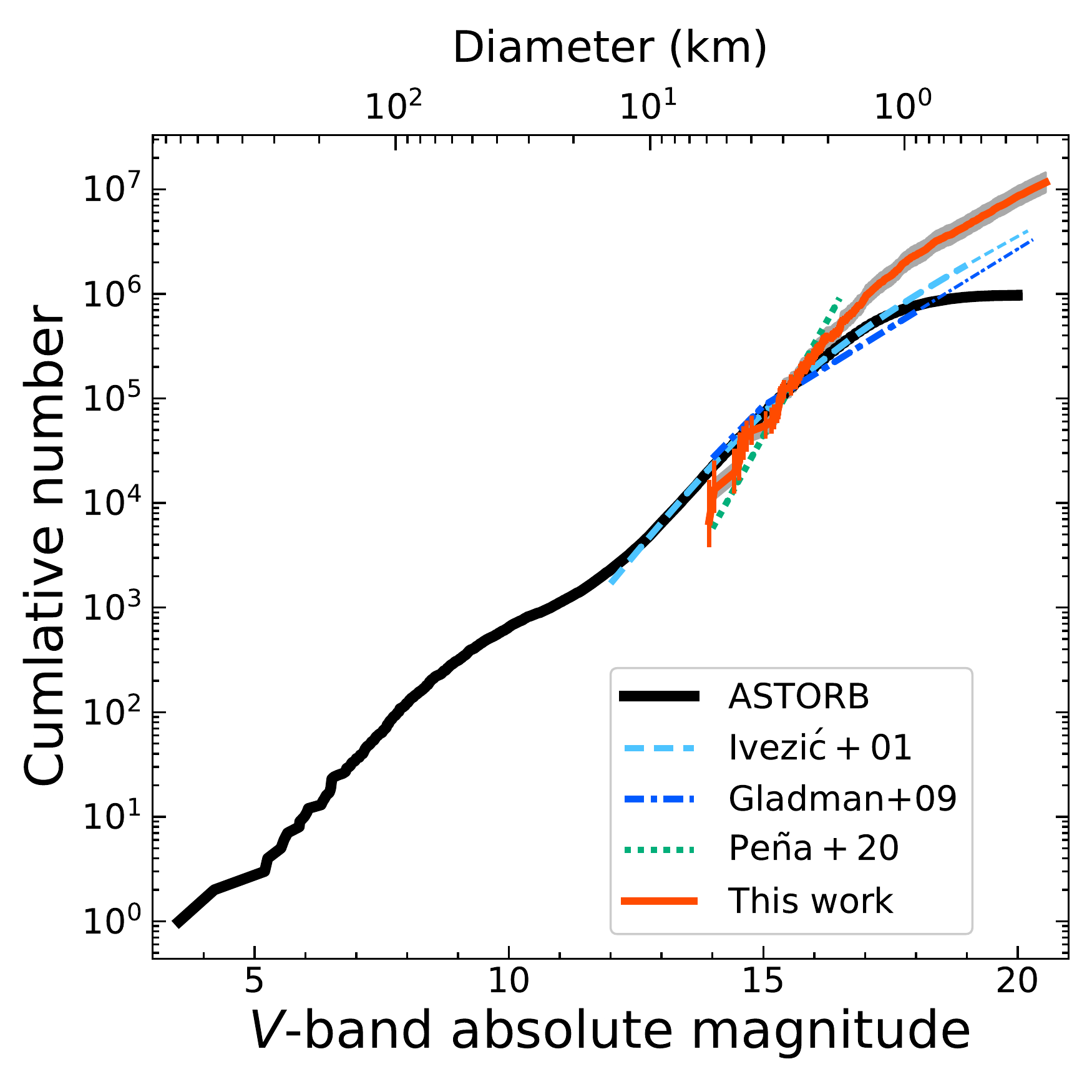}
	\end{center}

	\caption{Cumulative size distributions of MBAs obtained from previous studies and the present work. The red line shows the distribution derived from our sample (vertical lines show error bars). The black line is derived from known asteroids based on the ASTORB database \citep{b94}. The light-blue dashed line, the blue dot-dashed line, and the green dotted line are drawn with the model slopes obtained by \citet{i01}, \citet{g09}, and \citet{p20}, respectively, for reference (thin lines are their extrapolations). The ASTORB data are connected with our data at $H_V=15.4$~mag (the gray area shows the error range produced by the uncertainty of the cumulative number at the connection point), with \citet{i01} $H_V=13$~mag, and with both \citet{g09} and \citet{p20} at $H_V=15.5$~mag, respectively. Diameters shown on the upper horizontal axis are obtained by converting from absolute magnitudes under the assumption of an albedo $p=0.12$ \citep{u13}.}
	\label{fig:csd_all_othersurvey}
\end{figure}

We can roughly estimate the total number of MBAs by combining the size distributions of our work and previous works. In order to combine our results with those obtained in previous works, we converted the $r$-band magnitudes of our sample into the $V$-band magnitudes, assuming the average $m_V-m_r$ color of 0.25~mag based on the SDSS MOC4 data. \Figref{fig:csd_all_othersurvey} shows the combined cumulative size distribution of all MBAs. 
The red and the black lines show our sample and those derived from the ASTORB database \citep{b94}, respectively. For reference, we plotted the slopes obtained by \citet{i01}, \citet{g09}, and \citet{p20} with the light-blue dashed line, the blue dot-dashed line, and green dotted line, respectively (thin lines represent extrapolations of derived distributions). We combined the size distribution of our sample with that of ASTORB at $H_V=15.4$~mag, and the distributions obtained by \citet{i01}, \citet{g09}, and \citet{p20} are scaled so that they match that of ASTORB at $H_V=13$, 15.5, and $15.5$~mag, respectively, corresponding to the detection limit of each survey.

We found that the total number of MBAs brighter than $H_V=20$ mag is estimated to be $8.6 \times 10^{6}$, i.e., a couple of times as large as that obtained by extrapolating the results of \citet{i01} or \citet{g09}.
\edit1{We also found that the power-law slope of the absolute magnitude distribution at $H_V\simeq 13 - 17$~mag is nearly constant, which indicates that the break point found by the previous works may not be real. The potential reason for underestimating faint objects in the previous works is insufficient correction of detection bias. In our survey, the high sensitivity of the Subaru/HSC allowed high collection rate for faint objects. Also, we accurately corrected the number of detected asteroids considering the detection efficiency for each CCD image. It should be noted that the cumulative number of our sample at the combining point ($H_V=15.4$~mag) with ASTORB has uncertainties of $+2.6\times 10^{4}$  (25\%) and $-2.1\times 10^{4}$ (20\%), which produce uncertainties of $+2.1\times 10^{6}$ and $-1.7\times 10^{6}$ for the above estimation of the total number of MBAs.}

The fact that there are more small MBAs than previous estimate likely affect the estimation of generation rates of near earth asteroids (NEAs) and meteorites. Also, our newly-obtained accurate size distribution can be used as a constrain on theoretical studies of collisional and dynamical evolution of asteroids \citep[e.g.,][]{b05a,b05b,og05,db07}. In particular, the size distribution of small MBAs is important for comparative studies of asteroids based on analysis of crater size distribution and estimate of the crater retention ages for those asteroids visited by spacecraft \citep[e.g.,][]{b20}.

\begin{figure}[htb]
\begin{tabular}{cc}

\begin{minipage}{0.5\hsize}
	\begin{center}
		\includegraphics*[bb=0 0 558 500,scale=0.5]{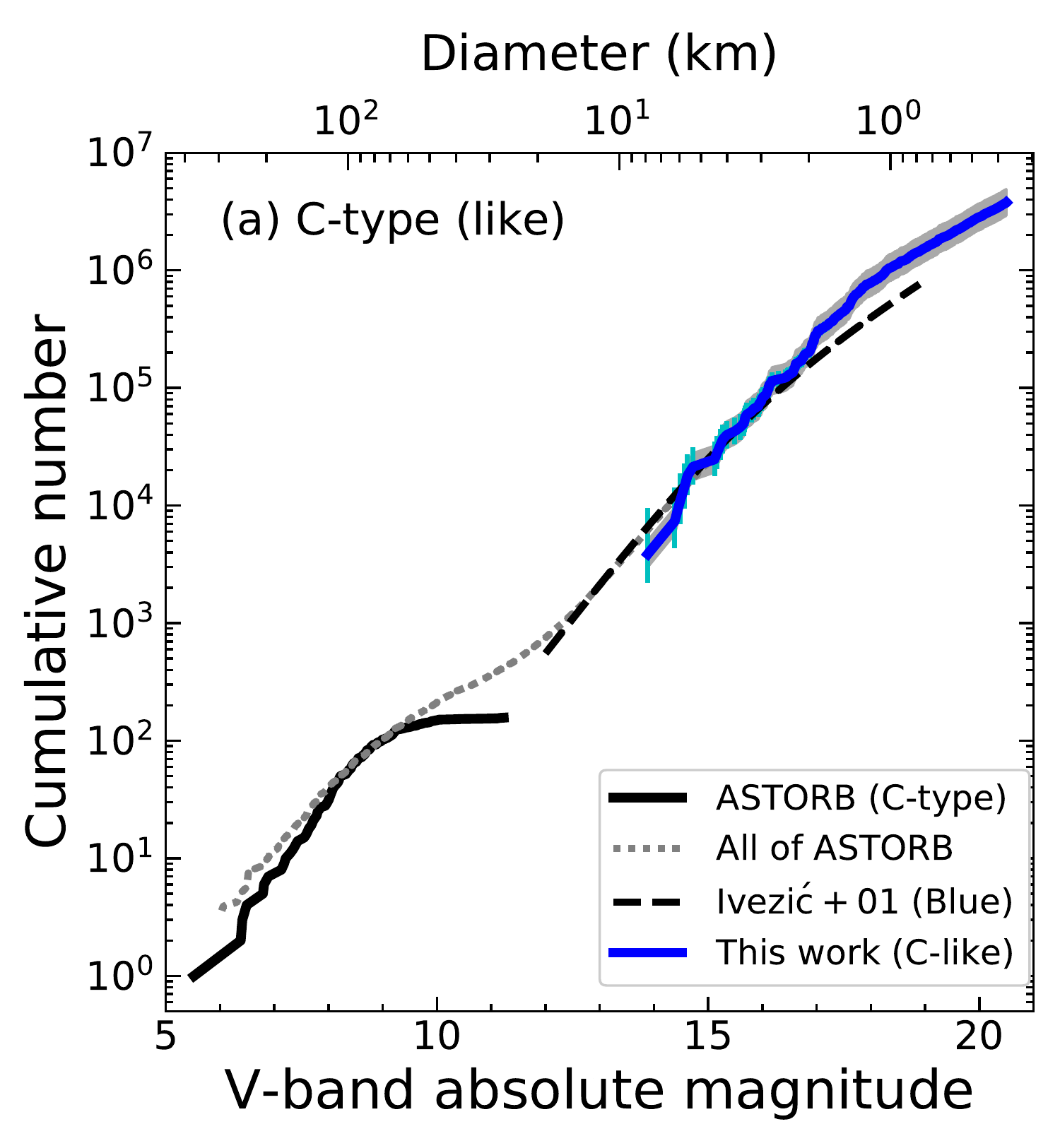}
	\end{center}
\end{minipage}&
	
\begin{minipage}{0.5\hsize}
	\begin{center}
		\includegraphics*[bb=0 0 558 500,scale=0.5]{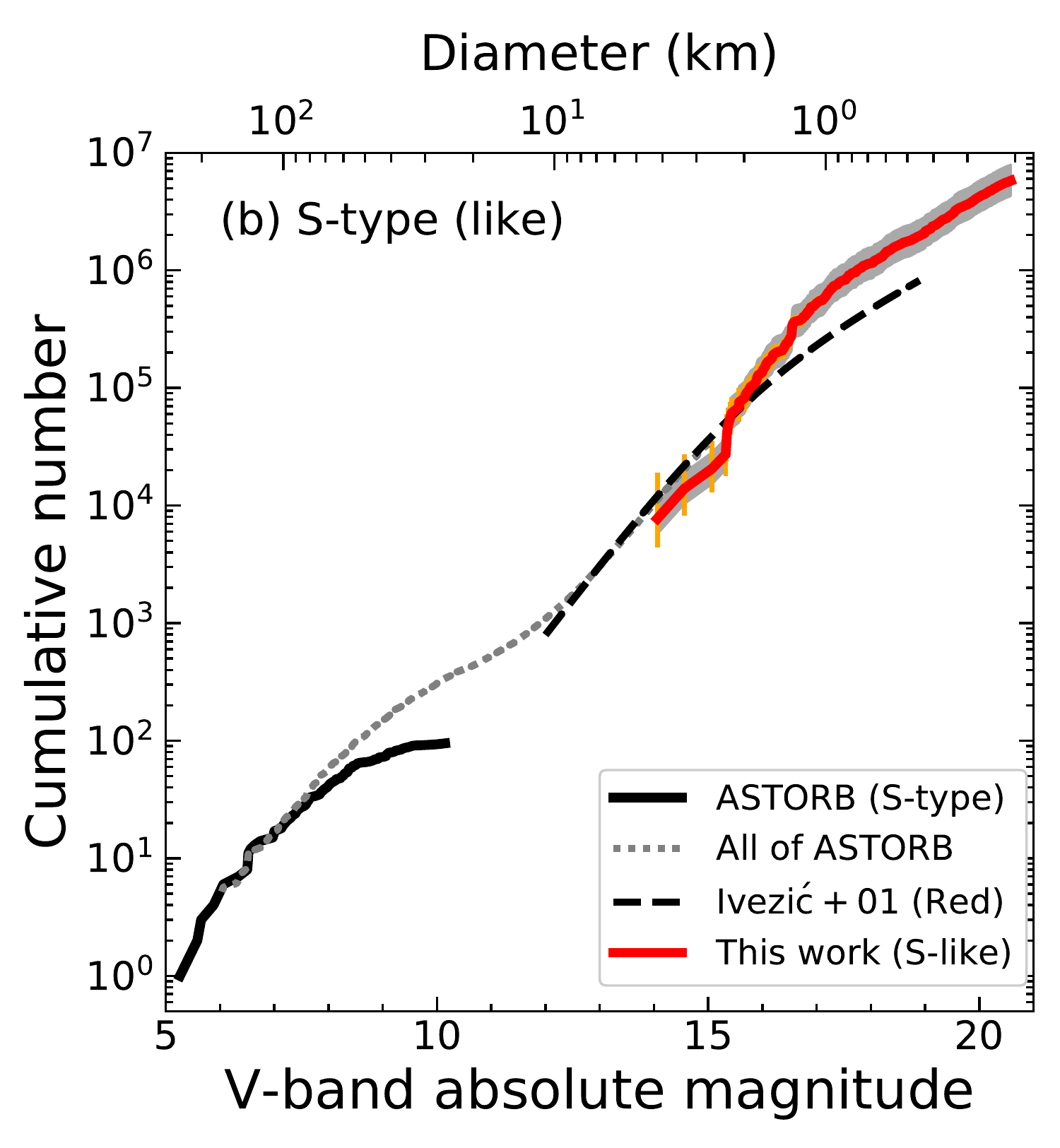}
	\end{center}
\end{minipage}

\end{tabular}
	\caption{Cumulative size distributions of C-type and S-type MBAs obtained from previous studies and the present work. The size distributions of known C-/S-type asteroids (black solid line) are connected with those of smaller sizes using the slopes of all MBAs in the ASTORB database (dotted line) at $H_V=9$~mag for C-type and $H_V=7$~mag for S-type, respectively. The size distribution of all ASTORB MBAs is connected with the blue/red asteroids of \citet{i01} (dashed line) at $H_V=13$~mag. The size distributions of our C-/S-like asteroids are connected with the blue/red asteroids of \citet{i01} at $H_V=15.5$~mag, respectively (the gray area shows the error range produced by the uncertainty of the cumulative number at the connection point). Error bars represent Poisson's errors. Diameters on the upper horizontal axis are obtained by converting from absolute magnitudes under the assumptions of an albedo (a) $p=0.07$ and (b) $p=0.21$, which are the mean values for the C-/S-type MBAs, respectively \citep{u13}.}
	\label{fig:csd_CS_othersurvey}
\end{figure}

\Figref{fig:csd_CS_othersurvey} shows the combined cumulative size distributions of C-type and S-type asteroids separately.
We combined the size distributions of our C-/S-like sample with the known C-/S-type asteroids (taken from the ASTORB database, black solid line) by interpolating with the slope of all known asteroids (dotted lines) and blue/red asteroids of \citet{i01} (dashed lines). Here, we converted the $r$-band magnitudes of our sample into the $V$-band magnitudes using the average $m_V-m_r$ colors of 0.20~mag for the C-type and 0.28~mag for the S-type asteroids, respectively. 
We can roughly estimate that the total number of C- and S-type MBAs larger than 400~m in diameter are about $3.5\times 10^6$ and $2.4\times 10^6$, respectively. Namely, the number of C-type asteroids is about 1.5 times as large as that of S-type asteroids for $D>400$~m.

\citet{yn07} found that the number ratio of C-like and S-like asteroids is 2.3 for $0.5<D<10$~km for the whole main-belt. Also, \citet{i01} reported that the number ratio of blue and red asteroids is 2.3 for $D>1$~km. Therefore, our result also shows a larger number of C-type asteroids than S-type, and is consistent with these previous works, although our result shows a smaller ratio.
\edit1{It should be noted that our ways of combining the size distributions have some uncertainties. First, the completeness limit of asteroids whose spectral type is obtained is uncertain, and actual size distributions in size range between the completeness limit of the known C-/S-type asteroids and bright end of the blue/red asteroids of \citet{i01} are unknown, which can primarily affect the total number. Second, the cumulative numbers of our samples at the combining points with ASTORB have uncertainties of $+1.3\times 10^{4}$ (30\%) and $-9.7\times 10^{3}$ (23\%) for C-like asteroids and $+2.2\times 10^{4}$ (37\%) and $-1.6\times 10^{4}$ (26\%) for S-like asteroids.} Also, our C-/S-like sample as well as the blue/red sample of \citet{i01} likely contain asteroids with types other than the C-type and S-type. More spectroscopic or multi-band photometric observation of asteroids for a wide range of sizes is desirable to more accurately estimate the total number of asteroids with each spectral type.

\section{CONCLUSIONS} \label{sec:conclusions}
We detected a large number of small MBAs in the $g$- and $r$-band images taken by the Hyper Suprime-Cam installed on the Subaru Telescope, and obtained an unbiased sample with a detection limit of $m_r=24.2$~mag.
We classified the sample by the $g-r$ color into two groups, bluish C-like and reddish S-like asteroids.

We found that the shapes of the size distributions of C-like and S-like asteroids agree well with each other for $0.4<D< 5$~km.
The similarity of their size distributions represents a similarity in the size dependence of impact strength of C-like and S-like asteroids for the size range from several hundred meters to several kilometers.
Assuming the asteroid population is under collision equilibrium, our results indicate that compositional difference hardly affects the size dependence of impact strength, at least for the above size range.
This size range is consistent with the size range where the observed distribution of asteroid rotation rates has an upper limit, called the spin-barrier. The existence of the spin-barrier can be interpreted as showing that most asteroids in this size range have rubble-pile structures, and our results are consistent with such a view.

By combining our results with previously known size distributions, we found that the total number of MBAs brighter than $H_V=20$~mag is a few times as large as that estimated by extrapolation of previous results. This would affect the estimation of the generation rates of NEAs and meteorites.

\acknowledgments
We are grateful to Takafumi Ootsubo, Eri Tatsumi, Tomokatsu Morota, and Hiroshi Kobayashi for constructive discussions. We also thank the anonymous referee for providing helpful comments and suggestions.
This work was supported by JSPS KAKENHI grant Nos. JP15H03716, JP16H04041, JP18K13607, JP20H04617, and JP21H00043.

\end{document}